\documentclass[acmsmall]{acmart}

\AtBeginDocument{%
  }

\setcopyright{cc}
\setcctype{by}
\acmDOI{10.1145/3832225}
\acmYear{2026}
\acmJournal{PACMSE}
\acmVolume{3}
\acmNumber{ISSTA}
\acmArticle{ISSTA134}
\acmMonth{10}
\acmSubmissionID{issta26main-p1266-p}
\received{2026-01-30}
\received[accepted]{2026-06-25}

\usepackage{color,soul}
\usepackage{multirow}
\usepackage{threeparttable}
\usepackage{booktabs}
\usepackage{subfigure}
\usepackage[many]{tcolorbox} % for summary box
\usepackage{fontawesome5} % for icons
\usepackage{colortbl}
\usepackage{enumitem}
\usepackage{threeparttable}
\usepackage{placeins}
\usepackage{xcolor}
\usepackage{hyperref}
\soulregister\ref7
\soulregister\cite7
\soulregister\pageref7

\usepackage[ruled,vlined,linesnumbered]{algorithm2e}
\SetAlgoVlined % 强制显示竖线和折线
\SetAlgoNoEnd  % 不显示 End while 等文字
\DontPrintSemicolon % 全局不打印分号
\SetKwInput{KwInput}{Input}
\SetKwInput{KwOutput}{Output}
\newtcolorbox{boxK}{
    top=2pt,
    bottom=2pt,
    left=2pt,
    right=2pt,
    boxrule = 0pt,
    toprule = 0pt, % top rule weight
}

\begin{document}

\title[CausalRepair: Bridging the Causality Gap in LLM-Based Automated Program Repair via Dual-Slicing]{CausalRepair: Bridging the Causality Gap in Large Language Model-Based Automated Program Repair via Dual-Slicing}

% \affiliation{%

% \affiliation{%

\author{Linhao Wu}
\email{wulinhao@mail.sdu.edu.cn}
\orcid{0009-0001-7624-156X}
\affiliation{%
   \department{School of Electronic and Computer Engineering}
   \department{Peking University Shenzhen Graduate School}
   \institution{Peking University}
   \city{Shenzhen}
   \country{China}
}

\author{Yizhou Chen}
\email{yizhouchen@stu.pku.edu.cn}
\orcid{0000-0003-1821-3170}
\affiliation{%
  \department{Key Laboratory of High Confidence Software Technologies (Peking University), MoE}
  \department{School of Computer Science}
  \institution{Peking University}
  \city{Beijing}
  \country{China}
}

\author{Zhen Yang}
\email{zhenyang@sdu.edu.cn}
\orcid{0000-0003-0670-4538}
\affiliation{%
   \department{School of Computer Science and Technology}
   \institution{Shandong University}
   \city{Qingdao}
   \country{China}
}

\author{Pengyu Xue}
\email{xuepengyu@mail.sdu.edu.cn}
\orcid{0009-0007-3395-9575}
\affiliation{%
  \department{Department of Computing}
  \institution{Hong Kong Polytechnic University}
  \city{Hong Kong}
  \country{China}
}

\author{Dan Hao}
\email{haodan@pku.edu.cn}
\orcid{0000-0001-8295-303X}
\affiliation{%
   \department{School of Electronic and Computer Engineering}
   \department{Peking University Shenzhen Graduate School}
   \institution{Peking University}
   \city{Shenzhen}
   \country{China}
}
\authornote{Corresponding author.}
% \renewcommand{\shortauthors}{Wu et al.}

\begin{abstract}
Automated Program Repair (APR) aims to automatically fix buggy programs. In recent years, with the rapid advancement of Large Language Models (LLMs), LLM-based APR techniques have achieved significant progress. Despite their potential, the effectiveness of LLMs relies heavily on the quality of the provided repair context. However, existing LLM-based APR approaches suffer from a causality gap when constructing such contexts.
Specifically, on the test side, existing methods struggle with test context ambiguity arising from noise interference or dependency absence;
meanwhile, on the source side, existing retrieval-augmented methods primarily rely on static analysis and inevitably introduce static over-approximation, resulting in contexts filled with unexecuted code and noise. Consequently, these contexts mislead LLMs, hindering them from identifying the true root cause and leading to incorrect fixes.

To bridge this gap, we introduce the concept of minimal causal context, defined as the essential set of dependencies required to explain a specific failure. Based on this, we propose CausalRepair, a novel conversation-driven APR framework that instantiates this concept through a synergistic dual-slicing strategy. Specifically, CausalRepair employs context-aware static slicing on the test side to purify test semantics, and utilizes execution-trace-based dynamic slicing on the source side to capture precise runtime dependencies. This constructs a high-quality context causally relevant to the bug, which filters out irrelevant code and guides the iterative repair process.
% Additionally, to further mitigate the overfitting problem, we employ patch augmentation to explore more diverse variations based on the generated plausible patches.
We evaluate CausalRepair on the widely used Defects4J (V1.2 and V2.0) and the latest Defects4J-Trans benchmarks. To ensure a fair comparison, we unify the backbone model as DeepSeek-V3 in all experiments. The results demonstrate that CausalRepair correctly fixes 313 bugs on Defects4J, significantly outperforming state-of-the-art approaches such as ReinFix and TSAPR, while reducing the average repair cost to \$0.029 per bug, achieving a dual optimization of effectiveness and efficiency.

\end{abstract}




\maketitle

\section{Introduction}

Automated Program Repair (APR) aims to automatically fix software bugs \cite{gazzola2018automatic, le2019automated, zhang2023survey}, serving as a critical technique for enhancing software reliability and development efficiency. 
% Early APR research primarily evolved through three traditional paradigms: heuristic-based \cite{jiang2018shaping, le2016history, wen2018context}, constraint-based \cite{le2017s3, mechtaev2016angelix, gao2021beyond}, and template-based \cite{martinez2016astor, liu2019avatar, liu2019tbar} techniques. Subsequently, learning-based \cite{chen2019sequencer, jiang2021cure, li2020dlfix, li2022dear, lutellier2020coconut, meng2022improving} approaches using Neural Machine Translation (NMT) models gained prominence. 
Early APR research primarily evolved from traditional techniques, i.e., heuristic-based \cite{jiang2018shaping, le2016history, wen2018context}, constraint-based \cite{le2017s3, mechtaev2016angelix, gao2021beyond}, and template-based \cite{martinez2016astor, liu2019avatar, liu2019tbar}, to learning-based approaches \cite{chen2019sequencer, jiang2021cure, li2020dlfix, li2022dear, lutellier2020coconut, meng2022improving}.
Recently, with the rapid advancement of Large Language Models (LLMs) and their swift adoption in software engineering practices \cite{nam2024using, xue2024automated, xue2025classeval, xue2025new}, LLM-based approaches \cite{xia2024automated, yin2024thinkrepair, kong2025contrastrepair, hu2025tsapr, zhang2025repair} have emerged as the state-of-the-art in APR, serving as the dominant paradigm in this field.

Despite the potential of LLMs, their repair effectiveness relies heavily on the quality of the repair context provided in the prompt \cite{parasaram2024fact, tian2024evaluating, yang2025survey}. 
To construct more effective contexts, existing works generally fall into two paradigms. The first is the \textit{execution-feedback driven} paradigm \cite{xia2024automated, hu2025tsapr, kong2025contrastrepair}, which dynamically evolves the repair context by incorporating error messages or pass-rate feedback from previous executions, providing feedback-augmented context that guides the LLM to progressively refine candidate patches and avoid repeating prior mistakes. The other is the \textit{retrieval-augmented context-aware} paradigm \cite{zhang2025repair,bouzenia2025repairagent, yin2024thinkrepair}, which enriches the repair context by introducing similar examples from the codebase and supplements the LLM with semantic dependencies and domain knowledge that are lacking in the local context. 
Although more context is provided, existing approaches still struggle to precisely align the repair context with the true cause of program failures. These contexts either do not match the semantics of the failures or contain a large amount of irrelevant noise. This issue is mainly manifested in the following two aspects.

% One is the \textit{execution-feedback driven} paradigm, such as ChatRepair \cite{xia2024automated}
% ContrastRepair \cite{kong2025contrastrepair}, 
% and TSAPR \cite{hu2025tsapr}. 
% These approaches leverage error messages or pass-rate feedback after test execution to guide the model through multi-round correction or search space exploration, leveraging validation signals to progressively refine candidate patches. 
% These approaches dynamically evolve the repair context by incorporating error messages or pass-rate feedback from previous executions, providing feedback-augmented context that guides the LLM to progressively refine candidate patches and avoid repeating prior mistakes.
% , and ThinkRepair \cite{yin2024thinkrepair}. 
% The other is the \textit{retrieval-augmented context-aware} paradigm, such as ReinFix \cite{zhang2025repair} and RepairAgent \cite{bouzenia2025repairagent}.
% These approaches enrich the repair context by introducing variable definitions, repair patterns, or similar examples from codebases via static analysis, vector retrieval, or agent tools, supplementing the LLM with semantic dependencies and domain knowledge absent from the local context.

% \textbf{Limitation1:}

% \textbf{Limitation2:}

% Novelty1: 我们提出了一个最小化因果上下文 。。。 解释，，， 怎么做的

% Novelty2:

% 分两段写limitation  然后分别写我们的novelty
% However, when balancing information sufficiency and noise interference, existing approaches struggle to establish a precise and dynamic dependency relationship between the software error manifestation and the root cause. This specifically manifests in two limitations: 

\textbf{Limitation 1: Test Context Ambiguity.} 
% To enable the LLM to understand test intent, some methods \cite{hu2025tsapr, zhang2025repair} provide the complete test function code in the prompt.
% On the test side, some approaches \cite{hu2025tsapr, zhang2025repair} include the complete test function code in the prompt to facilitate the LLM’s understanding of test intent. However, real-world test code often mixes multiple assertions targeting different scenarios. A recent study \cite{yang2025clarifying} indicates that mixing irrelevant assertions typically leads to complex logic, thereby blurring the semantic clarity of the test and impairing the LLM's use of the context. Other approaches \cite{xia2024automated, kong2025contrastrepair} provide only the failing assertion line or error messages, ignoring the complex context upon which the assertion depends (e.g., test fixture setup, contextual state manipulation), causing the LLM to fail in understanding the true intent of the test. 
On the test side, existing methods either offer overly complete test code, introducing multiple assertions and blurring the intent of failure; or only provide isolated failure assertions, ignoring the crucial execution context that the assertions rely on. In both cases, the model struggles to accurately understand the violated behavioral constraints, thereby weakening the guidance of test information for repair.

\textbf{Limitation 2: Source Context Over-approximation.} 
% On the source side, existing retrieval-augmented approaches \cite{zhang2025repair, bouzenia2025repairagent} primarily rely on static analysis to extract context dependencies. However, static analysis is inherently an over-approximation of program behavior, retrieving all possible dependencies and logical branches. When facing complex bugs, it is difficult for LLMs to distinguish which statements are executed and caused the bug, and which are irrelevant dead code. This static context, lacking runtime causality, forces the LLM to guess the connection between error phenomena and code logic, inducing hallucinated patches.
On the source side, existing approaches retrieve source code context using static analysis. Static analysis includes all possible dependencies and control-flow paths. Many of these paths are not executed when the failure occurs. Failure-related statements are mixed with irrelevant code. The LLM cannot tell which code actually causes the failure. This often leads to incorrect or hallucinated patches.

% 改一下这句话，把概念提出来
\textbf{Heuristically, effective APR depends on providing precise repair context that explicitly captures the causal relationship between test failures and the code statements that trigger them.} 
To address these limitations, we introduce the concept of \textbf{minimal causal context}, defined as the essential set of dependencies required to explain a specific failure. Building upon this concept, we propose CausalRepair, an LLM-based APR approach that instantiates this concept through a synergistic dual-end slicing strategy. To be specific, CausalRepair progressively refines the repair context through slicing on both the test side and the source code side, which is causally responsible for the observed test failure.
% To address the these limitations, we propose CausalRepair, an LLM based APR approach that centered on causal relationship. The core idea of CausalRepair is to explicitly restrict the repair context to information that is causally responsible for the observed test failure. To be specific, CausalRepair progressively refines the repair context through dual-end slicing on both the test side and the source code side. 
On the test side, CausalRepair performs dependency-aware slicing on failing tests to remove assertions and execution scenarios unrelated to the observed failure. This process retains only the violated verification logic together with its required initialization and state dependencies.
On the source code side, CausalRepair collects concrete runtime execution traces triggered by the refined test and conducts backward dynamic slicing. Unlike static analysis, dynamic slicing strictly follows the actual execution path and preserves only those statements that directly influence the failed assertion.
Finally, CausalRepair integrates the sliced verification logic from the test side with the dynamically sliced execution trace from the source code side to construct the \textbf{minimal causal context} that fully explains the failure. This context is then provided to the LLM to guide patch generation.

To evaluate the effectiveness of CausalRepair, we conduct extensive experiments on the widely used Defects4J (V1.2 and V2.0) benchmark and the latest Defects4J-Trans benchmark. {Specifically, we compare CausalRepair with 15 state-of-the-art baselines, covering template-based \cite{liu2019tbar}, learning-based \cite{xia2022less,jiang2021cure,ye2022neural,zhu2021syntax,ye2022selfapr,meng2023template,jiang2023knod}, and LLM-based \cite{xia2024automated,kong2025contrastrepair,yin2024thinkrepair,bouzenia2025repairagent,ye2025adverintent,hu2025tsapr,zhang2025repair} APR approaches.}
The experimental results show that CausalRepair outperforms prior APR tools. For example, CausalRepair can fix a total of 313 bugs on Defects4J, repairing 77 more bugs than the second-ranked ReinFix (236 bugs), representing a 32.6\% improvement. On Defects4J-Trans, CausalRepair can fix 289 bugs, achieving a 44.5\% improvement over state-of-the-art methods. {Additionally, benefiting from the high signal-to-noise ratio context provided by dual-slicing, CausalRepair significantly reduces token consumption while substantially improving repair rates, with an average cost of only \$0.029 per successfully fixed bug, which is significantly lower than or on par with current state-of-the-art methods, achieving a dual optimization of effectiveness and cost.}

In summary, the main contributions of this paper are as follows:

\setlist[itemize]{leftmargin=1em, labelsep=0.6em, itemsep=2pt}

% 三个贡献 1. 新概念 基于这个概念设计了技术 2. 评估 3. 开源代码
\begin{itemize}
    \item \textbf{Novel Perspective.} 
    We introduce the concept of minimal causal context, defined as the essential set of dependencies required to explain a failure. Building upon this, we propose CausalRepair, an LLM-based APR framework that instantiates this concept through a synergistic dual-slicing strategy. 
    \item \textbf{Extensive Evaluation.} We evaluate CausalRepair on the widely used Defects4J (V1.2 and V2.0) and the latest Defects4J-Trans benchmark datasets. {We compare CausalRepair with 15 state-of-the-art baselines.} 
    Experimental results show that CausalRepair can fix a total of 313 bugs on Defects4J, outperforming all studied baselines, while reducing repair costs.
    \item \textbf{Open Science.} To facilitate reproducibility and further research, we release the full implementation of CausalRepair, including the source code, experiment configurations, and experimental results. The project is publicly available at: \href{https://doi.org/10.5281/zenodo.20953899}{https://doi.org/10.5281/zenodo.20953899}.  
\end{itemize}

\section{Motivation}

This section utilizes a real-world case from Defects4J to demonstrate the limitations of existing LLM-based APR methods in context construction and thereby elucidate the necessity of CausalRepair's introduction of causal information.

% \vspace{-0.2in}

\subsection{Motivating Example}

\begin{figure*}[t]
    \centering
    \vspace{-0.1in}
    {\includegraphics[width=1.0\linewidth]{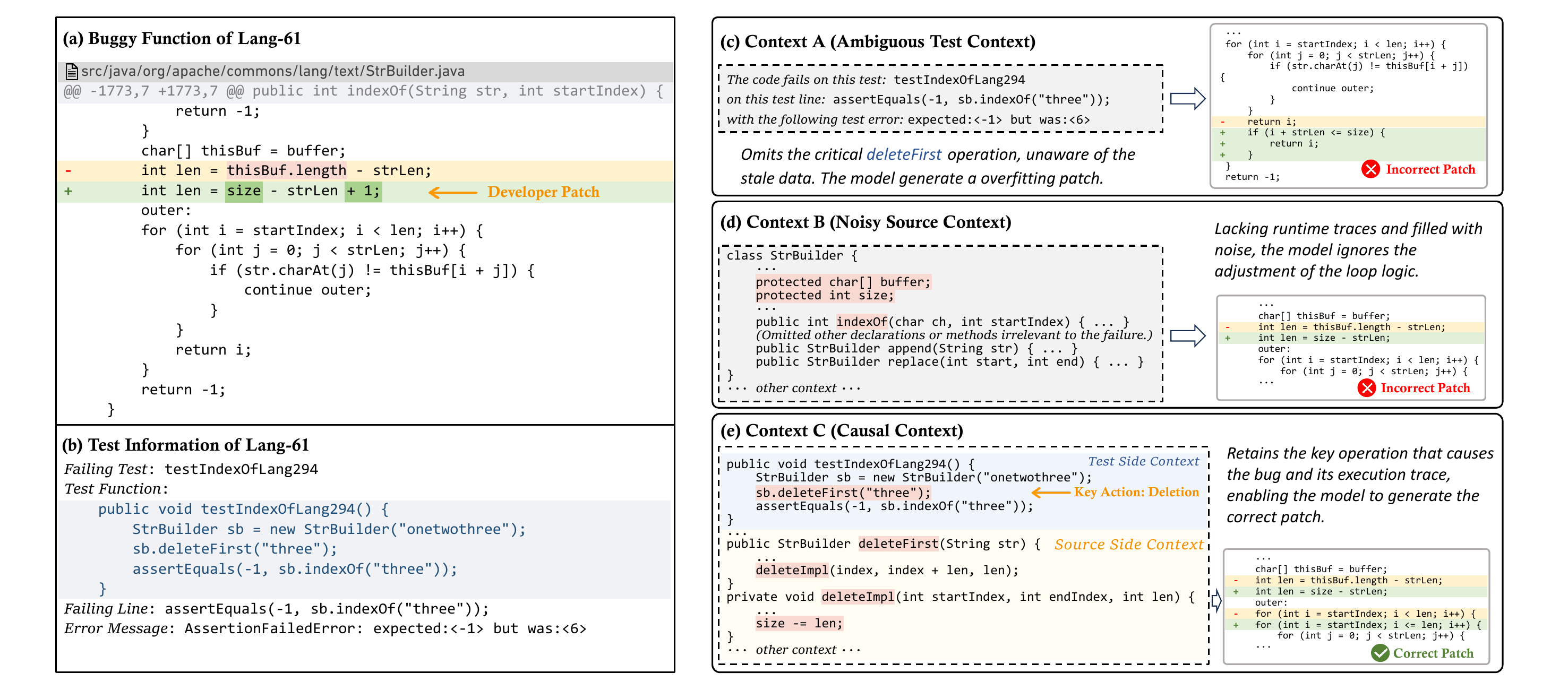}}
    \vspace{-0.3in}
    \caption{Motivation Example of CausalRepair.} 
    \label{fig:motivation}
    \vspace{-0.2in}
\end{figure*}

Figure~\ref{fig:motivation}-(a) and -(b) present the Lang-61 bug from the Apache Commons Lang project within Defects4J, along with its failure test information. This bug resides in the \texttt{indexOf} method of the \texttt{StrBuilder.java} file, which searches for the index of the first occurrence of a string starting at the given index. Specifically, \texttt{StrBuilder} internally maintains a character array \texttt{buffer} and a \texttt{size} variable representing the actual character length. The root cause of the bug lies in the yellow line, where \texttt{thisBuf.length} (i.e., the total capacity of the buffer) is erroneously used instead of \texttt{size} (the actual content length) to calculate the search boundary \texttt{len}. This bug is triggered by the test \texttt{testIndexOfLang294}. 
As shown in the test snippet, the program first initializes a \texttt{StringBuilder} object \texttt{sb} and then calls \texttt{sb.deleteFirst(...)} to remove a portion of the characters. This operation reduces \texttt{size}, but the \texttt{buffer} array retains the deleted characters. Because the buggy \texttt{indexOf} method uses the \texttt{buffer} capacity as the boundary, the search logic traverses beyond the valid region, incorrectly matching the deleted substring and resulting in an erroneous index return (Expected -1 but was 6). The developer's standard fix (as shown by the green line) not only replaces \texttt{thisBuf.length} with \texttt{size} but also adjusts the boundary calculation logic to \texttt{int len = size - strLen + 1;} to accommodate the \texttt{< len} loop condition. In practice, any fix that correctly delimits the valid content range (e.g., using \texttt{<=} in conjunction with \texttt{size - strLen}) is semantically equivalent and correct. Therefore, to repair this bug, the model must understand state dependency, meaning the valid search space should be determined by the dynamically changing \texttt{size}.

\paragraph{Context A: Ambiguous Test Context}
When the repair prompt includes only the failing assertion and error message, as shown in Figure~\ref{fig:motivation}-(c), the LLM observes that \texttt{indexOf("three")} unexpectedly returns a positive index but has no visibility into the preceding state mutation. Without observing the critical operation \texttt{deleteFirst("three")}, the model cannot infer the existence of stale data in the \texttt{buffer}. Consequently, it attributes the failure to local loop logic and generates an overfitting patch that satisfies the observed test but fails to address the underlying root cause.

\paragraph{Context B: Noisy Source Context}
In contrast, Figure~\ref{fig:motivation}-(d) illustrates a context constructed by augmenting the prompt with extensive static source code. Although this context exposes the declaration of \texttt{size}, it fails to convey its runtime evolution. The critical statement \texttt{size -= len}, which causally invalidates the original loop boundary, is buried among unrelated definitions. As a result, the LLM treats \texttt{size} as a static attribute without correctly adjusting the loop boundary semantics.

\paragraph{Context C: Causal Context}
As shown in Figure~\ref{fig:motivation}-(e), Context C explicitly preserves causally relevant information from both the test and the program execution. On the test side, the context explicitly preserves the mutation operation \texttt{sb.deleteFirst("three")} immediately preceding the failing assertion, signaling that the failure depends on a prior state change. On the program side, dynamic slicing traces this operation to the execution of \texttt{deleteImpl}, exposing the critical state update \texttt{size -= len}.
By jointly presenting the state-mutating action and its execution trace, Context C enables the LLM to establish a direct causal link: since \texttt{size} is reduced before \texttt{indexOf} is executed, the loop boundary must be constrained by \texttt{size}. This causal context allows the LLM to correctly repair the boundary computation, yielding a correct patch.

\subsection{Insight}
From this example, we draw a key insight: correct APR does not require more context, but \emph{causal context}. We propose CausalRepair, a causality-aware APR framework that constructs a \emph{minimal causal context}. CausalRepair selectively combines test-side slicing to retain failure-relevant actions with program-side dynamic slicing to capture execution traces that causally affect the buggy computation. By explicitly encoding cause–effect relationships, CausalRepair enables LLMs to reason about root causes and generate a correct patch.

\section{Approach}

\subsection{Framework Overview}

\FloatBarrier
\begin{figure*}[t]
    \centering
    \vspace{-0.1in}
    {\includegraphics[width=1.0\linewidth]{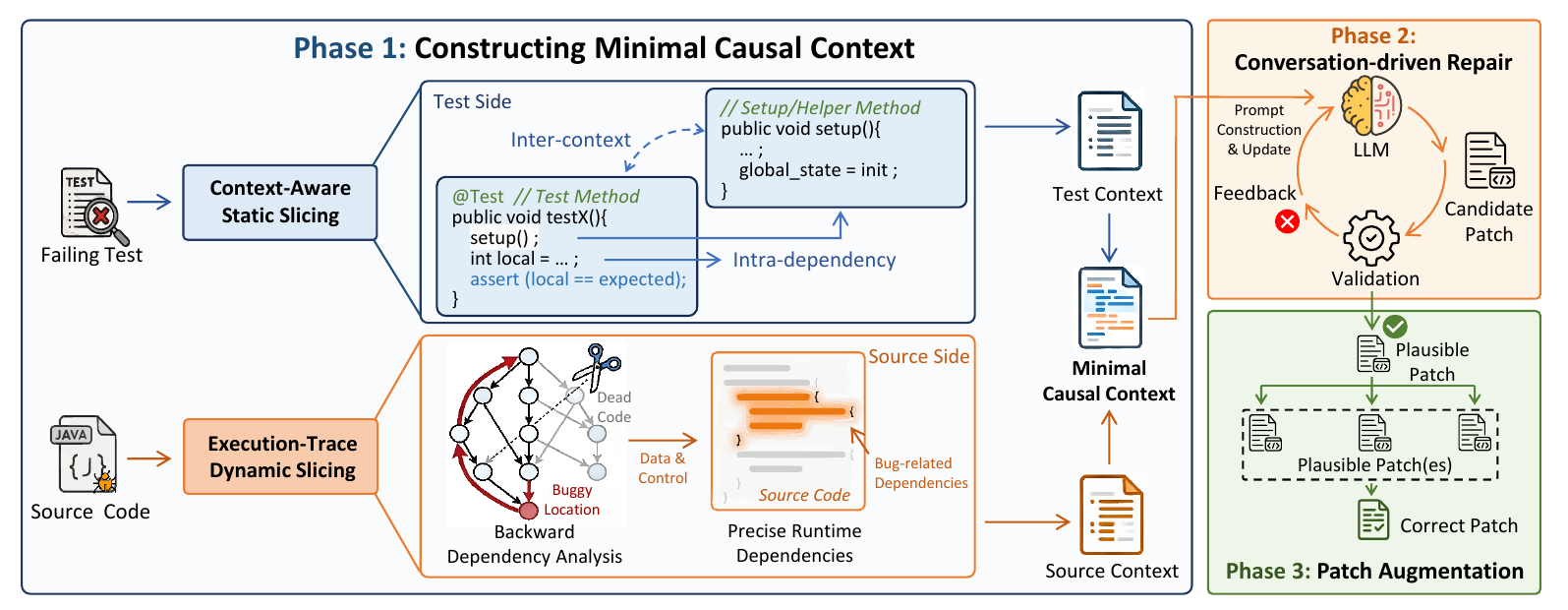}}
    \vspace{-0.3in}
    \caption{Overview of CausalRepair.}
    \label{fig:framework}
    \vspace{-0.1in}
\end{figure*}

In this section, we present the overall workflow of CausalRepair (as illustrated in Figure~\ref{fig:framework}), which consists of three distinct phases. 
In the Constructing Minimal Causal Context phase (Section~\ref{phase1}), a dual-slicing strategy extracts purified test context and dynamic execution paths to construct a refined prompt capturing the minimal causal dependencies of the bug.
In the Conversation-driven Iterative Repair phase (Section~\ref{phase2}), the repair process enters a multi-turn conversation in which the LLM iteratively refines the patch based on validation feedback to produce a plausible patch.
In the Patch Augmentation phase (Section~\ref{phase3}), the verified plausible patch serves as a reference to generate diverse variations, enhancing patch diversity and mitigating test overfitting.
% In the Constructing Minimal Causal Context phase, as detailed in Section~\ref{phase1}, a dual-slicing strategy is employed to extract both static test contexts and dynamic execution paths, with the goal of constructing a refined prompt that captures the minimal causal dependencies of the bug. In the Conversation-driven Iterative Repair phase, as detailed in Section~\ref{phase2}, the repair process enters a multi-turn conversation where the LLM iteratively refines the pacth based on feedback from validation failures, aiming to generate a plausible patch. In the Patch Augmentation phase, as detailed in Section~\ref{phase3}, the verified plausible patch serves as a reference exemplar to guide the generation of alternative variations, leveraging the LLM's generative capabilities to enhance patch diversity and mitigate test overfitting.

\subsection{Phase 1: Constructing Minimal Causal Context}
\label{phase1}

% Existing LLM-based APR approaches often face a dilemma when constructing the repair context: test context ambiguity, which stems from either noise interference in full tests or dependency absence when providing only the failing line. Additionally, relying solely on static analysis for source code retrieval suffers from static over-approximation, incorporating unexecuted dead code that hallucinates the LLM. To bridge this gap, we propose a dual-slicing strategy that constructs a minimal causal context by synergistically applying context-aware static slicing on the test side and execution-trace-based dynamic slicing on the source side.
Existing LLM-based APR approaches often face a dilemma in repair context construction: test context ambiguity, arising from noise in full tests or dependency absence when using only the failing line. Additionally, source code retrieval based solely on static analysis suffers from over-approximation, introducing unexecuted dead code that misleads the LLM. To bridge this gap, we propose a dual-slicing strategy that constructs a minimal causal context by applying context-aware static slicing on the test side and execution-trace-based dynamic slicing on the source side.

\subsubsection{Context-Aware Static Slicing on the Test Side}

The primary objective of this process is to resolve test context ambiguity by distilling the original test into a purified context that isolates the specific failure logic. 
{We achieve this by applying context-aware static slicing, which removes failure-irrelevant statements while preserving the structural and semantic cues needed for the LLM to understand the violated verification logic.}
To formalize this process, we define a test function as a sequence of statements $T = \langle s_1, s_2, ..., s_n \rangle$, where $s_{fail} \in T$ denotes the failing line that triggers the bug. Our goal is to derive a purified test context $\mathcal{C}_{test}$ by retaining statements causally related to $s_{fail}$ and filtering out irrelevant test scenarios. 
% {Static slicing is employed here to preserve structural and semantic cues, providing a coherent context for the LLM to interpret the violated verification logic.}
% {To achieve this goal, we employ context-aware static slicing, which removes failure-irrelevant statements while preserving the structural and semantic cues needed for the LLM to understand the violated verification logic.}
% that retains only the statements causally related to $s_{fail}$, removing noise from irrelevant test scenarios.

\textbf{Intra-procedural Dependency Analysis.} Inspired by Yang et al. \cite{yang2025clarifying}, we construct a System Dependence Graph (SDG) where edges represent dependencies between statements:
\begin{itemize}
    \item \textbf{Data Dependence ($s_i \xrightarrow{dd} s_j$):} Exists if a variable $v$ defined at $s_i$ is used at $s_j$, with no redefinition of $v$ in between.
    \item \textbf{Control Dependence ($s_i \xrightarrow{cd} s_j$):} Exists if the execution of $s_i$ (e.g., a predicate) determines whether $s_j$ is executed.
\end{itemize}
% Data Dependence ($s_i \xrightarrow{dd} s_j$): Exists if a variable $v$ defined at $s_i$ is used at $s_j$, with no redefinition of $v$ in between.
% Control Dependence ($s_i \xrightarrow{cd} s_j$): Exists if the execution of $s_i$ (e.g., a predicate) determines whether $s_j$ is executed.
We define the slicing criterion as $\kappa = \langle s_{fail}, V_{fail} \rangle$, where $V_{fail}$ is the set of variables referenced in $s_{fail}$. Based on the SDG, we compute the backward slice of $s_{fail}$ to obtain the intra-procedural slice set, denoted as $\mathcal{S}_{intra}$. This set contains all statements $s_i$ in $T$ that transitively affect $s_{fail}$:$$\mathcal{S}_{intra} = \{ s_i \in T \mid s_i \xrightarrow{*} s_{fail} \} \cup \{ s_{fail} \}$$where $\xrightarrow{*}$ denotes the transitive closure of Data Dependence ($\xrightarrow{dd}$) and Control Dependence ($\xrightarrow{cd}$). This step eliminates redundant lines targeting other test scenarios, clarifying the testing intent.

\textbf{Inter-procedural Context Extraction.} Merely retaining intra-procedural dependencies is insufficient, as test execution often relies on the external environment (e.g., fields, helper methods). To address the dependency absence problem, we extend the analysis to inter-procedural dependencies. Specifically, for any statement $s \in \mathcal{S}_{intra}$, if it invokes a method $m$ or references a field $f$, we recursively resolve their definitions into an external dependency set $\mathcal{E}$.
% \begin{itemize}
%     \item \textbf{Method Invocation:} If $s$ calls a helper method $m_{ext}$, we include the body of $m_{ext}$ into $\mathcal{E}$, recursively tracing calls up to a predefined depth.
%     \item \textbf{Fields and Setup:} If $s$ references a class field, we include its declaration and initialization logic (e.g., in setUp() or constructors) into $\mathcal{E}$.
% \end{itemize}
% Method Invocation: If $s$ calls a helper method $m_{ext}$, we include the body of $m_{ext}$ into $\mathcal{E}$, recursively tracing calls up to a predefined depth.
% Fields \& Setup: If $s$ references a class field, we include its declaration and initialization logic (e.g., in setUp() or constructors) into $\mathcal{E}$.
% Finally, we reconstruct the purified test context $\mathcal{C}_{test}$ by wrapping the merged dependencies $\mathcal{S}_{intra} \cup \mathcal{E}$ into a syntactically valid Java class structure. $\mathcal{C}_{test}$ serves as a precise driver for the subsequent dynamic analysis.

Finally, we merge $\mathcal{S}_{intra}$ with $\mathcal{E}$ to form $\mathcal{C}_{test}$. Additionally, we explicitly attach the concrete failing line $s_{fail}$ and the corresponding runtime error information to $\mathcal{C}_{test}$. Through this strategy, the resulting $\mathcal{C}_{test}$ constitutes a semantically purified and dependency-complete test fragment that not only makes the testing intent explicit but also provides a precise entry point for subsequent source-side dynamic slicing.

\subsubsection{Execution-Trace-Based Dynamic Slicing on the Source Side}

% While $\mathcal{C}_{test}$ clarifies the testing intent, retrieving relevant source code via static analysis (e.g., RAG) often introduces noise from unexecuted branches. To precisely isolate the root cause, we employ dynamic slicing to capture the runtime causality.

While $\mathcal{C}_{test}$ clarifies the testing intent, retrieving source code via static analysis inherently suffers from static over-approximation, often introducing noise from unexecuted statements. To address this problem, we employ execution-trace-based dynamic slicing to capture the precise runtime causality. This strategy rigorously filters out irrelevant logic, retaining only the statements that contributed to the specific failure.

\textbf{Trace Capture and Criterion Mapping.} We utilize \textit{Slicer4J} \cite{ahmed2021slicer4j}, a high-precision dynamic slicer, to capture the runtime behavior. We execute the buggy program $P$ driven by the purified test context $\mathcal{C}_{test}$. The execution trace is recorded as a sequence of events $\mathcal{T} = \langle e_1, e_2, ..., e_k \rangle$, where each $e_i$ represents an execution instance of a statement. To initiate slicing, we map the static failing line $s_{fail}$ to its corresponding runtime event. We define the dynamic slicing criterion as $\kappa_{dyn} = e_{target}$, where $e_{target} \in \mathcal{T}$ is the last execution instance corresponding to $s_{fail}$. This ensures the analysis is strictly scoped to the specific error manifestation.

\textbf{Backward Slicing on Dynamic Dependence Graph.} Starting from the criterion $e_{target}$, we construct a Dynamic Dependence Graph (DDG) over the execution trace $\mathcal{T}$. Unlike static analysis, dynamic dependencies are strictly defined by the actual interactions during execution:
\begin{itemize}
    \item \textbf{Dynamic Data Dependence:} Exists between events $e_i$ and $e_j$ ($i < j$) if $e_i$ defines a value consumed by $e_j$, and the value is not redefined between them.
    \item \textbf{Dynamic Control Dependence:} Exists if the outcome of $e_i$ (e.g., a branch predicate) directly determined the execution of $e_j$.
\end{itemize}
To obtain the final code context, we map the identified causal events back to the source code. Let $s$ denote a static statement in the buggy program. We compute the backward transitive closure of $e_{target}$ to extract the minimal set of causal statements $\mathcal{S}_{dyn}$:$$\mathcal{S}_{dyn} = \{ s \mid \exists e \in \mathcal{T}, \text{stmt}(e) = s \land e \xrightarrow{dyn*} e_{target} \}$$
Here, $\xrightarrow{dyn*}$ denotes the transitive dependence relation on the DDG, and $\text{stmt}(e)$ is the mapping function that returns the static statement $s$ corresponding to event $e$. This process physically filters out dead code and logic that, although executed, did not contribute to the specific failure.

% \textbf{Backward Slicing on Dynamic Dependence Graph.} Starting from $e_{target}$, we construct a Dynamic Dependence Graph (DDG) over $\mathcal{T}$. Unlike static dependencies, dynamic dependencies only consider interactions that actually occurred during execution:
% \begin{itemize}
%     \item \textbf{Dynamic Data Dependence:} Exists between $e_i$ and $e_j$ ($i < j$) if $e_i$ defines a variable $v$ used by $e_j$, and $v$ is not redefined between them.
%     \item \textbf{Dynamic Control Dependence:} Exists if the execution of $e_i$ (e.g., a branch predicate) directly determines the execution of $e_j$.
% \end{itemize}
% % Dynamic Data Dependence: Exists between $e_i$ and $e_j$ ($i < j$) if $e_i$ defines a variable $v$ used by $e_j$, and $v$ is not redefined between them.
% % Dynamic Control Dependence: Exists if the execution of $e_i$ (e.g., a branch predicate) directly determines the execution of $e_j$.
% We compute the backward transitive closure of $e_{target}$ to extract the minimal set of causal events:$$\mathcal{S}_{dyn} = \{ s \mid \exists e \in \mathcal{T}, \text{stmt}(e) = s \land e \xrightarrow{dyn*} e_{target} \}$$
% % $\mathcal{S}_{dyn}$ represents the minimal set of source statements that physically contributed to the failure, naturally filtering out irrelevant logic and dead code.
% $\text{stmt}(e)$ denotes the static source code statement corresponding to the event $e$, and $\mathcal{S}_{dyn}$ is the minimal set of such statements. This process physically filters out irrelevant logic and dead code.

\subsubsection{Structural Reconstruction of Causal Context}

Although $\mathcal{S}_{dyn}$ captures precise causal statements, they are physically fragmented. Directly feeding such discrete code lines to LLMs can lead to the "lost in the middle" \cite{liu2024lost} phenomenon due to the lack of structural context. To address this, we propose an Abstract Syntax Tree (AST)-based reconstruction strategy to restore semantic coherence, resulting in the final structured source context $\mathcal{C}_{source}$.

\textbf{Hierarchical Encapsulation.} We process the discrete statements in $\mathcal{S}_{dyn}$ to rebuild a valid code skeleton. First, we aggregate the statements by their originating files. For each file, we parse the statements to identify their enclosing Class and Method nodes. We then construct a Skeleton Tree that retains the original class definitions and method signatures, back-filling the sliced statements into their respective bodies. Code lines not selected by the slice are omitted, maximizing context compression while preserving the structural integrity required for LLM understanding.

\begin{figure*}[]
    \centering
    \vspace{-0.1in}
    {\includegraphics[width=1.0\linewidth]{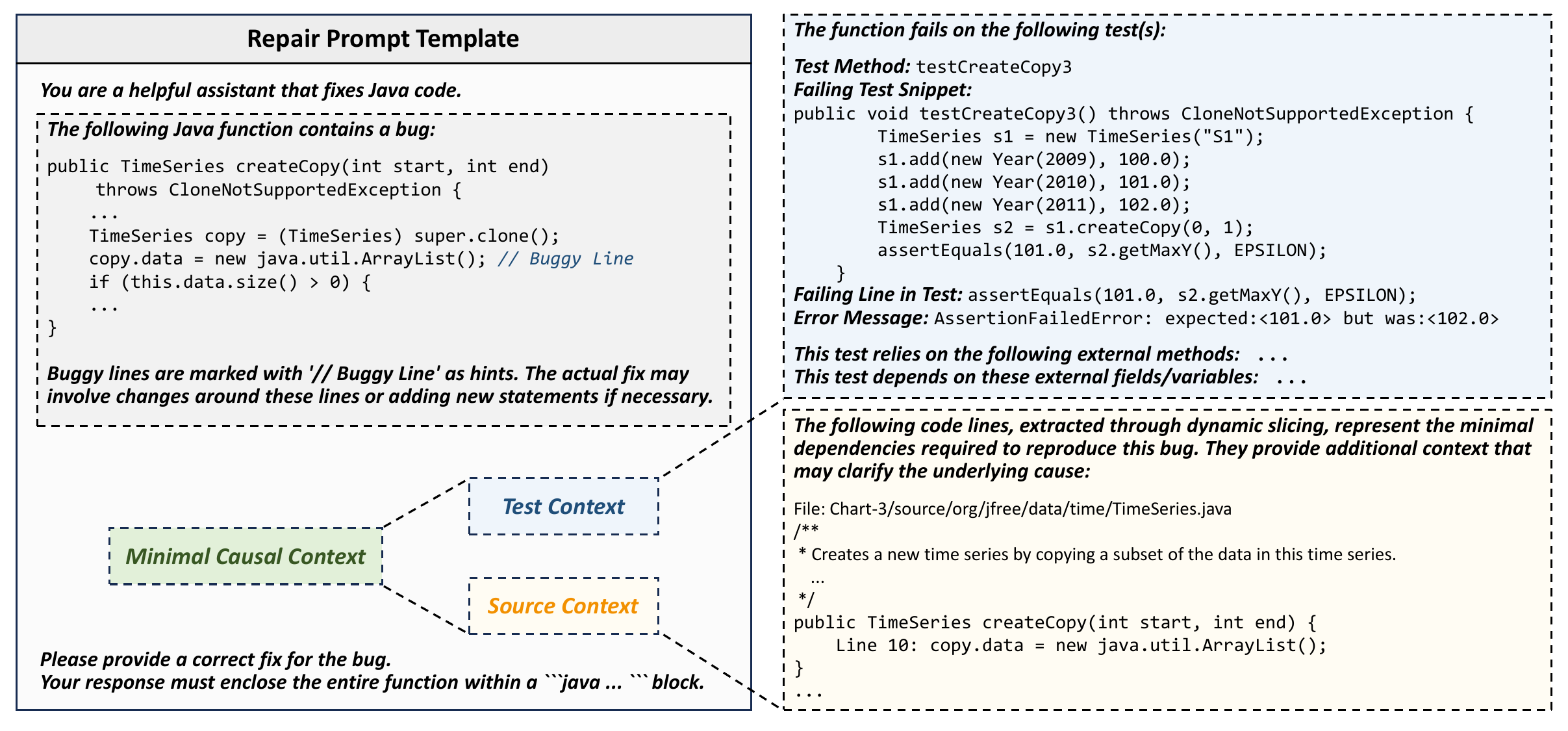}}
    \vspace{-0.3in}
    \caption{Illustration of the Prompt Construction.} 
    \label{fig:prompt}
    \vspace{-0.1in}
\end{figure*}

\textbf{Semantic Anchor Injection.} To further enhance the LLM's spatial and functional reasoning, we inject multi-dimensional meta-data into $\mathcal{C}_{source}$:
\begin{itemize}
    \item \textbf{File Path Headers:} To clarify cross-module boundaries, we inject the file path (e.g., \texttt{// File: .../TimeSeries.java}) as a header at the beginning of the code block for each distinct file. 
    % This header appears only once per file, grouping all relevant methods and classes from that file under a unified context.
    \item \textbf{Intent Preservation:} We explicitly preserve \textit{Javadoc} and comments associated with the retained classes and methods. These natural language descriptions provide high-level intent, bridging the gap between code implementation and semantic purpose.
    \item \textbf{Position-Aware Markers:} Relative line numbers (e.g., \texttt{// Line 10:}) are appended to each sliced statement to help the LLM establish spatial awareness for patch generation.
    % Thto help the LLM establish spatial awareness for patch generation.is helps the LLM establish spatial awareness, enabling it to generate patches that align with the original line numbers.
\end{itemize}
Finally, $\mathcal{C}_{source}$ is combined with the test context $\mathcal{C}_{test}$ to form the final prompt. Figure~\ref{fig:prompt} illustrates the constructed prompt using a specific example (Chart-3 bug). This structured input eliminates noise while preserving the complete chain of evidence and program context, enabling the LLM to perform precise causal reasoning.

\subsection{Phase 2: Conversation-Driven Iterative Repair} 
\label{phase2}

Following the construction of the causal context in Phase 1, we incorporate this context into a conversation-driven APR process \cite{xia2024automated, kong2025contrastrepair}. 
% CausalRepair proceeds to the conversation-driven iterative repair phase. 
Instead of simple repeated sampling or one-shot generation, this phase adopts a debugging process analogous to human developers, employing a closed-loop “generate–validate–feedback” mechanism that enables the LLM to progressively refine patches through multi-turn conversation. By leveraging negative feedback from previous iterations, the model incrementally corrects errors rather than restarting the repair process from scratch.

Algorithm~\ref{algo:causalrepair} outlines the overall repair process of CausalRepair. Given a buggy function and its test suite, the algorithm aims to generate plausible patch(es) that pass all tests. 
% In each iteration, a prompt is constructed (Line 11 or Line 13) and fed to the LLM to produce a candidate patch (Line 14). Subsequently, the patch is injected into the project, compiled, and validated using the test suite (Line 15). 
{In each iteration, CausalRepair constructs a prompt (Line 11 or Line 13) and queries the LLM to generate one candidate patch at a time (Line 14). The generated patch is then immediately injected into the project, compiled, and validated using the test suite (Line 15).
The validation result determines the subsequent flow of the repair process. 
If the patch passes all tests (including both triggering and regression tests), it is identified as a plausible patch, at which point the iterative repair process terminates and CausalRepair proceeds to the patch augmentation phase (Phase 3).}
% If the patch passes all tests (including both triggering and regression tests), it is identified as a plausible patch and the process proceeds to the patch augmentation phase (Phase 3). 
However, if the patch fails validation, CausalRepair does not simply discard it; instead, it initiates the feedback mechanism. Specifically, to guide the LLM towards effective self-correction, we design an iterative prompting strategy incorporating negative feedback. This prompt consists of two key components: (1) the incorrect candidate patch generated in the previous iteration, and (2) the failing test(s) and the specific error messages (e.g., compilation errors or runtime exceptions) captured during validation. We then request the LLM to analyze the failure and provide a corrected function.

\newcommand{\algcomment}[1]{%
  \footnotesize\ttfamily\textcolor{gray}{#1}%
}
\SetCommentSty{algcomment}

\begin{algorithm}[!t]
\small
  \KwInput{$Func$: the buggy function, $\mathcal{T}$: the test suite,
  $m$: the maximum number of repair attempts, $n$: the maximum number of iterative interactions per attempt
  }
  \KwOutput{$PFuncs$: the plausible patch(es)}
  
   % Phase 1: Dual-Slicing Context Construction
   $T_{fail} := \texttt{IdentifyFail}(\mathcal{T}, Func)$; \tcp*{locate the failing test function}
   $C_{test} := \texttt{StaticSlice}(T_{fail})$; \tcp*{Phase 1: purify test context}\label{algo:staticslice}
   $C_{source} := \texttt{DynamicSlice}(Func, C_{test})$; \tcp*{Phase 1: extract source context}\label{algo:dynamicslice}
   
   $iter_1:=0$;  \tcp*{initiate the number of repair attempts}
   
   \While{$iter_1 < m$}{\tcp*[r]{restart repair from original function} \label{algo:whilerestart}
       $H := \emptyset$; \tcp*{initialize conversation history}
       $tmp := Func$\;
       $iter_2:=0$;  \tcp*{initiate the number of iterative interactions per attempt}
       
       \While{$iter_2 < n$}{\tcp*{continuous repair from previous patch}\label{algo:whilecontinue}
           % Prompt Construction
           \If{$iter_2 = 0$}{
               $prompt := \texttt{ConstructInit}(tmp, C_{test}, C_{source})$; \tcp*{construct causal context}
           }
           \Else{
               $prompt := \texttt{ConstructFeed}(V, H)$; \tcp*{construct feedback from validation and history}
           }
           
           % LLM Interaction
           $Func' := \texttt{LLM}(prompt)$; \tcp*{Phase 2: query LLM for repair}\label{algo:invokellm}
           $V := \texttt{Validate}(Func', \mathcal{T})$; \tcp*{validate the patched function} \label{algo:validate}
           
           % Evaluation & Augmentation
           \If{$V = \texttt{True}$}{
               $PFuncs:=\texttt{PatchAug}(Func')$; \tcp*{Phase 3: generate alternative patches} \label{algo:patchaug}
               \textbf{return} $PFuncs$;\label{algo:correctpatch} \tcp*{return plausible patches}
           }
           
           % Update Context for Next Turn
           $H := \texttt{UpdateHistory}(H, prompt, Func')$; \tcp*{accumulate context}
           $tmp := Func'$; \tcp*{update current function}
           $iter_2 = iter_2 + 1$;
       }
       $iter_1 = iter_1 + 1$;
   }
   \Return $\emptyset$;  \tcp*{fail to repair}

\caption{CausalRepair}
\label{algo:causalrepair}
\end{algorithm}

This process continues until a plausible patch is generated or the preset maximum number of interactions is reached. By combining the causal context extracted in Phase 1 with the iterative validation feedback, CausalRepair can effectively converge to correct repairs through multi-turn interactions, enhancing the overall repair effectiveness.

\subsection{Phase 3: Patch Augmentation} 
\label{phase3}

While Phase 2 successfully produces a plausible patch that passes all test cases, this does not mark the end of the repair task. Existing APR research \cite{jiang2023knod, jiang2021cure, xia2024automated} indicates that due to test suite incompleteness, passing tests alone cannot guarantee semantic correctness; patches are often merely overfitting solutions tailored to specific test inputs. To mitigate this problem, following prior work \cite{xia2024automated, kong2025contrastrepair}, CausalRepair introduces the patch augmentation phase.

In this phase, the plausible patch generated in Phase~2 is treated as a high-value reference rather than the final output. The intuition behind this is that even if a plausible patch is not fully correct, it often captures essential logic or structural patterns required to fix the bug. 
Therefore, rather than generating from scratch, it is more effective to explore the solution space surrounding this existing solution. Specifically, CausalRepair instructs the LLM to produce alternative implementations that are logically similar to the reference patch. These variations are then validated against the test suite and the validated candidates are returned as outputs. 
This process effectively increases patch diversity, thereby improving the likelihood that the final results presented to developers contain the correct patch, and mitigating the potential bias introduced by relying on a single plausible patch.

\section{Experimental Setup}

\subsection{Research Questions}
We evaluate CausalRepair on the following research questions (RQs):
\begin{itemize}
    \item \textbf{RQ1 (Repair Effectiveness): How does the repair effectiveness of CausalRepair compare to state-of-the-art APR  techniques?} \\ This question evaluates the repair effectiveness of CausalRepair in Java bug repair and compares its performance to existing state-of-the-art APR baselines.
    \item \textbf{RQ2 (Repair Scenarios): How does CausalRepair perform across different repair scenarios?} \\ This question investigates CausalRepair’s performance in three distinct repair scenarios (i.e., single-function, single-hunk, and single-line bugs).
    \item \textbf{RQ3 (Ablation Study): How does each design choice of CausalRepair affect its repair effectiveness?} \\ This question investigates the individual contribution of key components and further evaluates the impact of key hyperparameter settings on the overall repair effectiveness.
    \item {\textbf{RQ4 (Data Contamination Analysis): Is CausalRepair's effectiveness affected by potential training-data contamination?} \\ 
    Considering the potential data leakage risk caused by LLMs, this question evaluates whether CausalRepair maintains its repair effectiveness on a contamination-reduced benchmark that is less likely to appear in the model's training corpus.}
    % Considering the potential data leakage risk caused by LLMs, this question assesses CausalRepair’s generalization ability by evaluating its performance on recent real-world bugs that are guaranteed to be unseen during training.
    \item \textbf{RQ5 (Cost Analysis): How does the cost of CausalRepair compare to existing LLM-based methods?} \\ Considering the use of LLMs may lead to higher repair costs, this question evaluates the time and monetary overhead of CausalRepair to assess its practical feasibility.
\end{itemize}

\subsection{Benchmarks}

\begin{itemize}
    \item \textbf{Defects4J.}
    To evaluate the repair effectiveness, we adopt the widely used benchmark Defects4J~\cite{just2014defects4j}. Following prior APR studies~\cite{xia2023automated, xue2024exploring, xia2024automated, yin2024thinkrepair, zhang2025repair, hu2025tsapr}, we divide Defects4J into two versions, V1.2 and V2.0. Defects4J V1.2 consists of 391 bugs in 6 different Java projects, while Defects4J V2.0 consists of 438 new bugs across 9 additional projects. Consistent with prior studies~\cite{zhu2021syntax, xia2024automated, yin2024thinkrepair, zhang2025repair, hu2025tsapr}, we further categorize Defects4J into three repair scenarios: single-function (SF), single-hunk (SH), and single-line (SL). Specifically, single-function refers to bugs where the fix is confined within a single method, whereas single-hunk and single-line represent finer-grained scenarios requiring modifications to a continuous code block and a single line, respectively. Note that the single-line category is a subset of the single-hunk category, and the single-hunk category is a subset of the single-function category. The statistics of each scenario in Defects4J is presented in Table~\ref{tab:details_d4j}.
    \item \textbf{Defects4J-Trans.}
    {To assess potential data-leakage effects, we further conduct experiments on a recent contamination-reduced benchmark, namely Defects4J-Trans~\cite{li2025evaluating}.} Defects4J-Trans consists of 483 bugs, which are generated by applying semantic-preserving transformations to the 483 single-function bugs in the Defects4J benchmark (versions 1.2 and 2.0). Similar to Defects4J, Defects4J-Trans can also be categorized into three repair scenarios (i.e., SF, SH, and SL). The statistics for each scenario in Defects4J-Trans are also presented in Table~\ref{tab:details_d4j}. Since Defects4J-Trans was first released in October 2024 (i.e., later than the cutoff date of DeepSeek-V3’s training corpus), it serves as a benchmark that is less prone to data contamination for providing a contamination-reduced setting for evaluating repair performance.
\end{itemize}

\begin{table}[h]
\scriptsize
\vspace{-0.1in}
\caption{Statistics of studied benchmarks.}
\vspace{-0.1in}
\label{tab:details_d4j} 
% \resizebox{1\columnwidth}{!}{
\begin{tabular}{@{}ccccc|cccc@{}}
\toprule
\multirow{2}{*}{\textbf{Benchmarks}} & \multicolumn{4}{c|}{\textbf{V1.2}}                   & \multicolumn{4}{c}{\textbf{V2.0}}                    \\ \cmidrule(l){2-9} 
                                    & \# Total Bugs & \# SF Bugs & \# SH Bugs & \# SL Bugs & \# Total Bugs & \# SF Bugs & \# SH Bugs & \# SL Bugs \\ \midrule
\textbf{Defects4J}                  & 391           & 255        & 154        & 80         & 438           & 228        & 159        & 78         \\
\textbf{Defects4J-Trans}            & 255           & 255        & 154        & 80         & 228           & 228        & 159        & 78         \\ \bottomrule
\end{tabular}
% }
\vspace{-0.2in}
\end{table}

\subsection{Baselines}

To evaluate the effectiveness, {we compare CausalRepair against 15 state-of-the-art baselines across different categories,} including seven learning-based APR methods (i.e., AlphaRepair \cite{xia2022less}, CURE \cite{jiang2021cure}, RewardRepair \cite{ye2022neural}, Recoder \cite{zhu2021syntax}, SelfAPR \cite{ye2022selfapr}, TENURE \cite{meng2023template}, and KNOD \cite{jiang2023knod}), one template-based traditional APR method (i.e., TBar \cite{liu2019tbar}), {and seven LLM-based APR methods (i.e., ChatRepair \cite{xia2024automated}, ContrastRepair \cite{kong2025contrastrepair}, ThinkRepair \cite{yin2024thinkrepair}, RepairAgent \cite{bouzenia2025repairagent}, AdverIntent-Agent \cite{ye2025adverintent}, TSAPR \cite{hu2025tsapr}, and ReinFix \cite{zhang2025repair}).} 
Additionally, following prior work \cite{xia2024automated, kong2025contrastrepair, yin2024thinkrepair}, we construct an LLM-based baseline named BaseDeepSeek, which directly uses DeepSeek-V3 for repair without providing additional feedback and slicing information, serving as a foundational comparison. Following the common practice in the APR community \cite{xia2023plastic, xia2022less, xia2024automated, zhu2021syntax, kong2025contrastrepair, yin2024thinkrepair}, we reuse the reported results from previous studies for non-LLM baselines, while all LLM-based baselines are reproduced under a unified model to ensure fair comparison.

% \vspace{-0.1in}

\subsection{Evaluation Metrics}

Following previous work~\cite{jiang2023knod, jiang2021cure, kong2025contrastrepair, yin2024thinkrepair, zhang2025repair, hu2025tsapr}, we consider two widely used metrics to evaluate the effectiveness of both CausalRepair and baselines:

\begin{itemize}
    \item \textbf{Plausible Fixes:}
    Refers to the number of bugs which can pass all the test cases after fixing, without any additional verification.
    \item \textbf{Correct Fixes:}
    % Measures the ability of the repair tool to produce accurate patches. 
    Counts the number of bugs that have been properly repaired based on a manual review of the plausible patches generated by each tool.
\end{itemize}

\subsection{Implementation}

For the experiments, we utilize DeepSeek-V3 and GPT-4o as the backbone models for CausalRepair, using the APIs provided by SiliconFlow \cite{SiliconF82:online} and OpenAI \cite{ModelOpe18:online}, respectively. Crucially, to ensure a fair comparison and eliminate the confounding variable of model capability, we move beyond the reported results in prior work. {Instead, we establish a unified evaluation benchmark by reproducing all compared LLM-based APR methods (i.e., ChatRepair \cite{xia2024automated}, ContrastRepair \cite{kong2025contrastrepair}, ThinkRepair \cite{yin2024thinkrepair}, RepairAgent \cite{bouzenia2025repairagent}, AdverIntent-Agent \cite{ye2025adverintent}, TSAPR \cite{hu2025tsapr}, and ReinFix \cite{zhang2025repair}) using the same DeepSeek-V3 backbone.} During reproduction, apart from replacing the backbone model with DeepSeek-V3, we strictly adhere to the default architectural designs and hyperparameter settings described in their original papers, ensuring that any performance divergence is solely attributable to the methodology itself rather than the underlying model. For CausalRepair, following previous work \cite{xia2024automated, kong2025contrastrepair, yin2024thinkrepair, zhang2025repair}, we set the sampling temperature to 1.0 to obtain a diverse set of potential patches. In terms of fault localization (FL), to avoid additional biases introduced by FL tools, we also follow recent works \cite{xia2024automated, kong2025contrastrepair, yin2024thinkrepair, hu2025tsapr, zhang2025repair} and adopt the perfect fault localization setting. Regarding the repair budget, we set the number of repair attempts ($m$) to 5, with the maximum number of iterative interactions per attempt ($n$) set to 3. Additionally, we set the number of queries for the patch augmentation stage to 10. Therefore, the total maximum patch space explored by CausalRepair for each bug is 25 (i.e., $5 \times 3 + 10$). This budget is lower than the number of patches generated by state-of-the-art methods such as TSAPR \cite{hu2025tsapr} and ReinFix \cite{zhang2025repair}. We allocate a time budget of 5 minutes for dynamic slicing; if a timeout or crash occurs, the dynamic context is omitted. All experimental evaluations are conducted on one server running Ubuntu 20.04.

% \vspace{-0.1in}
\section{Experimental Results}

\subsection{RQ1: Repair Effectiveness}

% 1. 实验设置，怎么去看表  2. 总分：定性的说明结论 然后定量的分析结果 （解释）
% To answer RQ1, we compare the number of plausible fixes and correct fixes generated by CausalRepair against the baseline methods on Defects4J V1.2 and V2.0. Table~\ref{tab:result_d4j} presents the comparison results, including the overall results and the number of project-level bug fixes, of CausalRepair and baseline methods on the Defects4J benchmark.
To answer RQ1, we compare the numbers of plausible and correct fixes produced by CausalRepair and the baseline methods on Defects4J V1.2 and V2.0. The results, including overall performance and the number of project-level bug fixes, are presented in Table~\ref{tab:result_d4j}.

\begin{table*}[ht]
% \vspace{-0.1in}
\scriptsize
\setlength{\tabcolsep}{2.5pt}
\begin{threeparttable}
% \vspace{-0.1in}
\caption{Repair results (correct fixes / plausible fixes) for CausalRepair and baselines on Defects4J. 
}
\vspace{-0.1in}
\label{tab:result_d4j} 
% \resizebox{\textwidth}{!}{
\begin{tabular}{@{}ccc|cccccc|cc@{}}
\toprule
Category                                                                            & APR Method                                  & Patch Size                 & Chart                         & Closure                       & Lang                          & Math                          & Mockito                       & Time                         & \#Total (D4J V1.2)                       & \#Total (D4J V2.0)                       \\ \midrule
Template-based                                                                      & TBar \cite{liu2019tbar}                                        & -                          & 11/-                          & 16/-                          & 13/-                          & 22/-                          & 3/-                           & 3/-                          & 68/95                                    & 8/25                                     \\ \midrule
                                                                                    & AlphaRepair \cite{xia2022less}                                 & 5000                       & 9/-                           & 23/-                          & 13/-                          & 21/-                          & 5/-                           & 3/-                          & 74/109                                   & 36/-                                     \\
                                                                                    & CURE \cite{jiang2021cure}                                        & 5000                       & 10/-                          & 14/-                          & 9/-                           & 19/-                          & 4/-                           & 1/-                          & 57/-                                     & 19/-                                     \\
                                                                                    & RewardRepair \cite{ye2022neural}                                & 200                        & 5/-                           & 15/-                          & 7/-                           & 19/-                          & 3/-                           & 1/-                          & 50/-                                     & 25/-                                     \\
                                                                                    & Recoder \cite{zhu2021syntax}                                     & 100                        & 11/-                          & 22/-                          & 9/-                           & 22/-                          & 1/-                           & 2/-                          & 67/-                                     & 18/-                                     \\
                                                                                    & SelfAPR \cite{ye2022selfapr}                                     & 150                        & 7/-                           & 20/-                          & 10/-                          & 22/-                          & 3/-                           & 3/-                          & 65/79                                    & 45/51                                    \\
                                                                                    & TENURE \cite{meng2023template}                                    & 500                        & 7/-                           & 26/-                          & 16/-                          & 22/-                          & 4/-                           & 4/-                          & 79/-                                    & 50/-                                    \\
\multirow{-7}{*}{Learning-based}                                                   & KNOD \cite{jiang2023knod}                                        & 1000                       & 10/11                         & 23/29                         & 11/13                         & 20/25                         & 5/5                           & 2/2                          & 71/85                                    & 50/85                                    \\ \midrule
                                                                                    & BaseDeepSeek                                & 25                         & 9/13                          & 11/13                         & 20/31                         & 31/43                         & 5/6                           & 2/7                          & 78/113                                   & 79/96                                    \\
                                                                                    & ChatRepair \cite{xia2024automated}                                  & 500                        & 13/15                         & 11/13                         & 26/40                         & 38/64                         & 10/13                         & 5/11                         & 103/156                                  & 42/53                                    \\
                                                                                    & ContrastRepair \cite{kong2025contrastrepair}                              & 160                        & 13/16                         & 41/59                         & 31/39                         & 35/51                         & 8/8                           & 5/9                          & 133/182                                  & 48/66                                    \\
                                                                                    & ThinkRepair \cite{yin2024thinkrepair}                                & 125                        & 11/14                         & 22/32                         & 20/24                         & 27/40                         & 5/8                           & 2/4                          & 87/122                                   & 93/129                                   \\
                                                                                    & RepairAgent \cite{bouzenia2025repairagent}                                 & 160                        & 10/12                         & 23/26                         & 15/16                         & 31/32                         & 4/5                           & 2/2                          & 85/93                                    & 87/94                                    \\
                                                                                     &  AdverIntent-Agent \cite{ye2025adverintent}                                 &  36                        &  10/11                         &  20/23                         &  16/18                         &  31/34                         &  5/5                           &  2/2                          &  84/93                                    &  87/96                                    \\
                                                                                    & TSAPR \cite{hu2025tsapr}                                       & 32                         & 12/13                         & 25/41                         & 20/24                         & 36/48                         & 10/11                         & 5/9                          & 108/146                                  & 116/150                                  \\
                                                                                    & ReinFix \cite{zhang2025repair}                                     & 45                         & 11/15                         & 34/54                         & 26/36                         & 37/58                         & 6/11                          & 4/7                          & 118/181                                  & 118/171                                  \\
\multirow{-8}{*}{\begin{tabular}[c]{@{}c@{}}LLM-based\\ (DeepSeek-V3)\end{tabular}} & \cellcolor[HTML]{EFEFEF}CausalRepair        & \cellcolor[HTML]{EFEFEF}25 & \cellcolor[HTML]{EFEFEF}14/16 & \cellcolor[HTML]{EFEFEF}46/53 & \cellcolor[HTML]{EFEFEF}35/38 & \cellcolor[HTML]{EFEFEF}51/61 & \cellcolor[HTML]{EFEFEF}10/12 & \cellcolor[HTML]{EFEFEF}6/11 & \cellcolor[HTML]{EFEFEF}\textbf{162/191} & \cellcolor[HTML]{EFEFEF}\textbf{151/174} \\ \midrule
                                                                                    & ReinFix \cite{zhang2025repair}                             & 45                         & 14/-                          & 38/-                          & 25/-                          & 34/-                          & 9/-                           & 4/-                          & 124/-                                    & 130/-                                    \\
\multirow{-2}{*}{\begin{tabular}[c]{@{}c@{}}LLM-based\\ (GPT-4o)\end{tabular}}      & \cellcolor[HTML]{EFEFEF}CausalRepair & \cellcolor[HTML]{EFEFEF}25 & \cellcolor[HTML]{EFEFEF}12/15 & \cellcolor[HTML]{EFEFEF}46/57 & \cellcolor[HTML]{EFEFEF}34/37 & \cellcolor[HTML]{EFEFEF}41/52 & \cellcolor[HTML]{EFEFEF}11/13 & \cellcolor[HTML]{EFEFEF}8/10 & \cellcolor[HTML]{EFEFEF}\textbf{152/184} & \cellcolor[HTML]{EFEFEF}\textbf{145/175} \\ \bottomrule
\end{tabular}
% }

\begin{tablenotes}
\footnotesize
\item ``-'': indicates no results reported in the original work.
\end{tablenotes}

\end{threeparttable}

\end{table*}

\textbf{Overall Effectiveness.} {As shown in Table~\ref{tab:result_d4j}, CausalRepair achieves the best performance among all 15 baseline methods.} Note that to ensure a fair comparison, all LLM-based APR methods are reproduced under a unified experimental setting, where DeepSeek-V3 is used as the backbone model. Specifically, when instantiated with DeepSeek-V3, CausalRepair correctly fixes a total of 313 bugs.
% , with 162 bugs in Defects4J V1.2 and 151 bugs Defects4J in V2.0. 
Compared to the best-performing baselines reproduced under the same model, CausalRepair fixes 77 and 89 more bugs than ReinFix (236 bugs) and TSAPR (224 bugs), respectively. Additionally, compared to BaseDeepSeek (157 bugs) which directly uses the model for repair, CausalRepair fixes 156 more bugs, demonstrating its effectiveness in improving repair performance. Furthermore, CausalRepair demonstrates consistent effectiveness across different dataset versions. On Defects4J V1.2, CausalRepair fixes 162 bugs, surpassing the previous state-of-the-art methods ContrastRepair (133 bugs) and ReinFix (118 bugs). On the more challenging Defects4J V2.0, which contains more complex logic errors and longer functions, CausalRepair maintains its advantage by fixing 151 bugs. In contrast, ReinFix and TSAPR only fix 118 and 116 bugs, respectively. These results demonstrate the superiority of CausalRepair.

\textbf{Impact of Backbone Models.} To evaluate the generalizability of CausalRepair across different LLMs, we further implement CausalRepair using GPT-4o and compare it with the GPT-4o-based results reported by ReinFix \cite{zhang2025repair}. As shown at the bottom of Table~\ref{tab:result_d4j}, CausalRepair with GPT-4o correctly fixes a total of 297 bugs (152 in Defects4J V1.2 and 145 in Defects4J V2.0), which exceeds the performance reported by ReinFix with GPT-4o (254 bugs). This result indicates that the effectiveness of CausalRepair is not limited to specific model architectures. Instead, providing accurate causal dependency information can consistently enhance the repair capability of LLMs.

\begin{figure}[htbp]
\centering
    \subfigure[Venn on Defects4J V1.2]{
        \includegraphics[width=0.35\columnwidth]{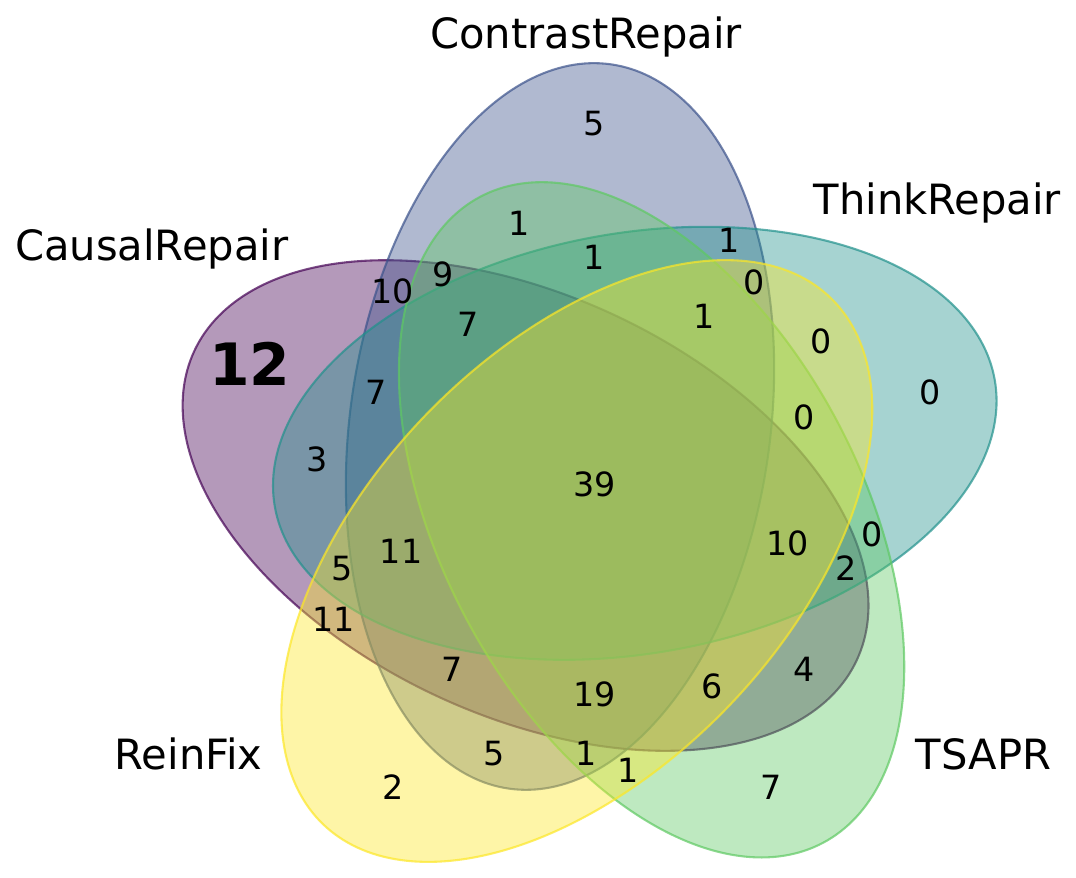}
        \label{venn_d4j12}
    }
    \subfigure[Venn on Defects4J V2.0] {
        \includegraphics[width=0.35\columnwidth]{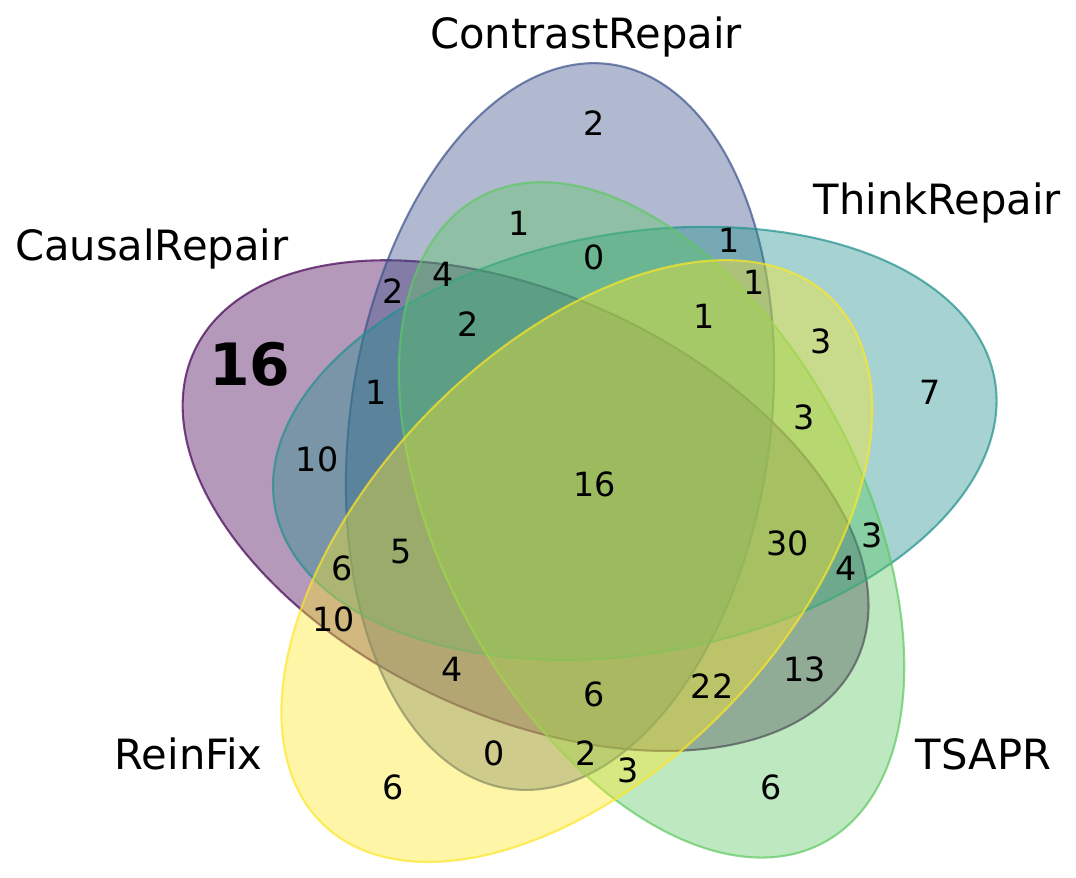}
        \label{venn_d4j20}
    }
    \vspace{-0.15in}
    \caption{Bug Fix Venn Diagram on Defects4J (CausalRepair, ContrastRepair, ThinkRepair, TSAPR, ReinFix)}
    
    \label{d4j_venn}
\end{figure}

\textbf{Overlap and Unique Fixes Analysis.} We further analyze the overlap of fixed bugs among the top-performing methods, including ReinFix, TSAPR, ContrastRepair, and ThinkRepair. Figure~\ref{d4j_venn} presents the Venn diagrams of unique and shared fixes on Defects4J V1.2 and Defects4J V2.0. 
Figure~\ref{venn_d4j12} and Figure~\ref{venn_d4j20} show that 39 and 16 bugs on Defects4J V1.2 and V2.0, respectively, can be successfully repaired by all these methods. This overlap indicates that all these approaches are highly effective and exhibit considerable similarity in their repair capabilities. This similarity can be attributed to the fact that all these methods are implemented on the same backbone model.
Despite that, CausalRepair is still able to uniquely fix 12 and 16 bugs on Defects4J V1.2 and V2.0, respectively, which are not fixed by any other methods. These results suggest that CausalRepair serves as a complement to existing work and highlights the unique advantage of CausalRepair in capturing fine-grained causal dependencies.

\vspace{-0.1in}

% \textbf{Case Study.} To better demonstrate the advancement of CausalRepair, we further demonstrate a unique fix produced from CausalRepair, as shown in Figure \ref{fig:case}.  Lang-18 is a function-level bug from Defects4J V1.2 that cannot be fixed by simply replacing one or several buggy lines. Fixing this bug involves changes at multiple locations within the function, which poses significant challenges to APR tools and no baselines can fix it. The bug resides in the \texttt{init} method of \texttt{FastDateFormat}, where a missing conditional causes year patterns of length 3 (e.g., ``yyy'') to be incorrectly assigned to the two-digit year field, producing formatting errors. While the developer resolves this by introducing nested conditional checks to isolate \texttt{tokenLen == 2}, CausalRepair generates a syntactically distinct but semantically equivalent fix. This demonstrates CausalRepair’s ability to produce correct and concise solutions  in addressing complex logic errors.

\begin{figure*}[h]
    \centering
    % \vspace{-0.1in}
    {\includegraphics[width=1.0\linewidth]{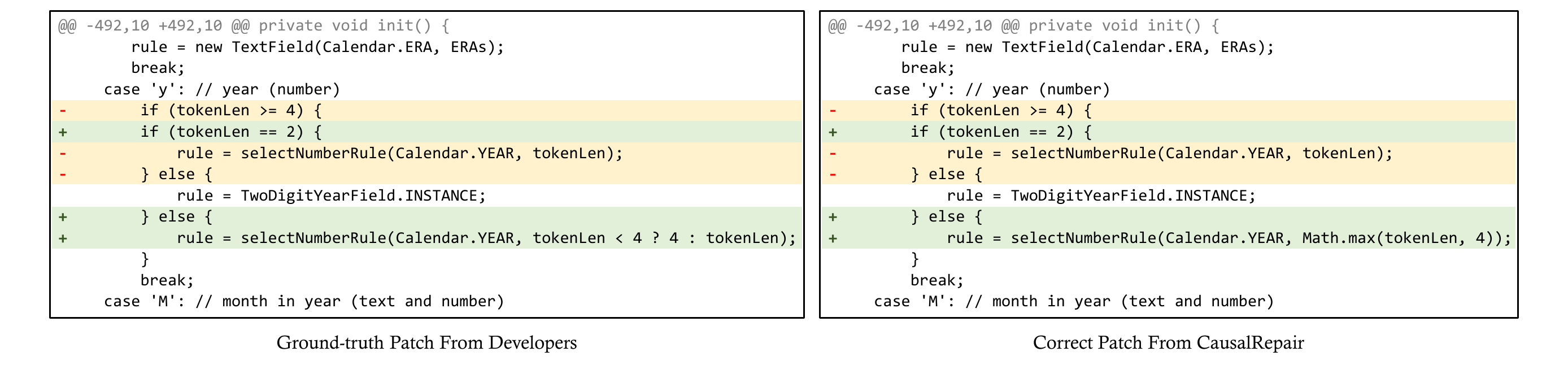}}
    \vspace{-0.3in}
    \caption{Unique Fix Generated by CausalRepair for Lang-18.}
    \label{fig:case}
    \vspace{-0.2in}
\end{figure*}

\begin{figure*}[h]
    \centering
    % \vspace{-0.1in}
    {\includegraphics[width=1.0\linewidth]{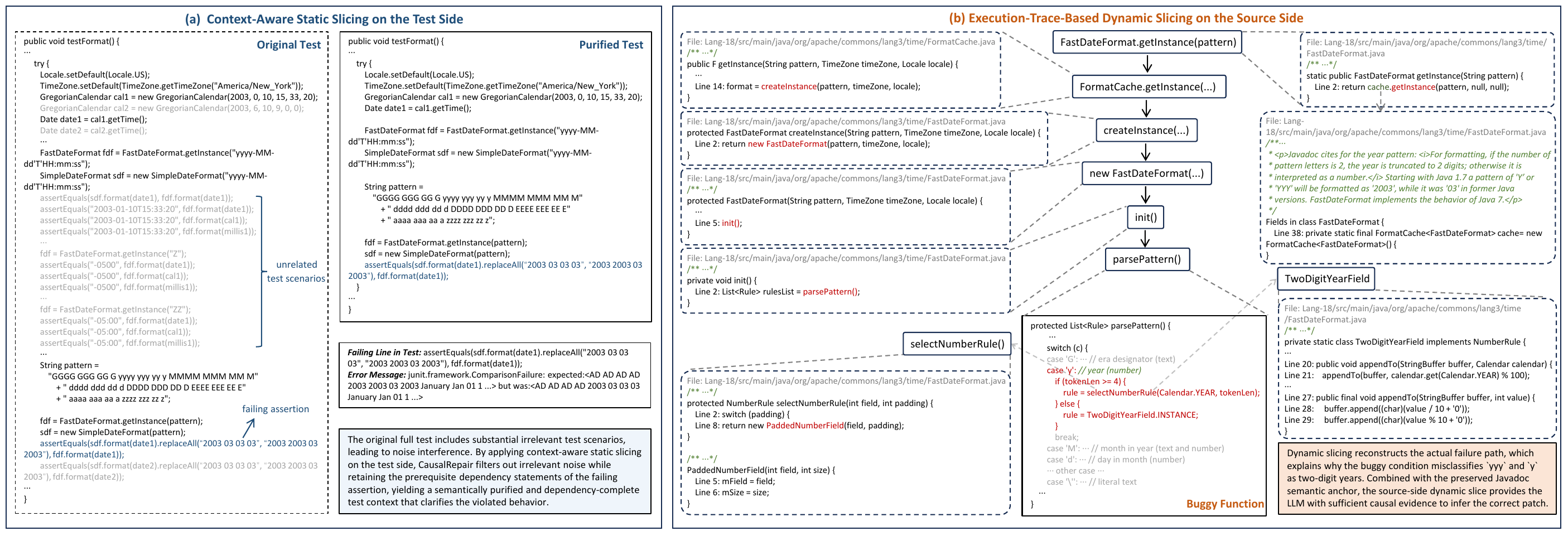}}
    \vspace{-0.3in}
    \caption{Dual-Slicing Context for the Lang-18 Case Study.}
    \label{fig:slicing_case}
    \vspace{-0.1in}
\end{figure*}

{\textbf{Case Study.} To provide qualitative evidence on how dual-slicing context helps CausalRepair, we analyze Lang-18, a unique bug fixed only by CausalRepair, as shown in Figure~\ref{fig:case} and~\ref{fig:slicing_case}. Lang-18 is a function-level bug in \texttt{FastDateFormat}, triggered by a compound date pattern containing year tokens of different lengths, i.e., \texttt{yyyy yyy yy y}. In the original test, the failing assertion is surrounded by many unrelated formatting checks. As shown in Figure~\ref{fig:slicing_case}-(a), test-side slicing removes these irrelevant scenarios while preserving the dependencies of the failing assertion, making the violated behavior explicit: only the two-letter year token should be formatted as a two-digit year. Figure~\ref{fig:slicing_case}-(b) shows the source-side dynamic slice, which reconstructs the actual runtime path from \texttt{FastDateFormat.getInstance(pattern)} to \texttt{parsePattern()} and the \texttt{case 'y'} branch. This slice exposes the root cause: the buggy condition routes all year tokens with \texttt{tokenLen < 4} to \texttt{TwoDigitYearField}. Since \texttt{TwoDigitYearField} truncates the year using \texttt{Calendar.YEAR \% 100} and emits exactly two digits, \texttt{yyy} and \texttt{y} are incorrectly formatted as \texttt{03} instead of \texttt{2003}. Together with the preserved \textit{Javadoc} semantic anchor, which states that only two-letter year patterns should be truncated, the dual-slicing context provides a complete causal chain from the failing test pattern to the incorrect rule selection. Based on this evidence, CausalRepair infers the general repair rule that only \texttt{tokenLen == 2} should use \texttt{TwoDigitYearField}, producing a patch that is syntactically different from but semantically equivalent to the developer patch, as shown in Figure~\ref{fig:case}.
}

\FloatBarrier

\FloatBarrier

\subsection{RQ2: Repair Scenarios}

To answer RQ2, we categorize the bugs in Defects4J into three repair scenarios based on granularity: single-function (SF), single-hunk (SH), and single-line (SL). The single-function scenario is generally considered the most challenging because it requires the model to understand the logic of the entire function and handle potential multiple modifications, whereas the single-line scenario is relatively simple. Table~\ref{tab:result_three_repair_scenarios} details the number of plausible and correct fixes achieved by different LLM-based methods under these scenarios.
Note that the results for ChatRepair and ContrastRepair in the SF and SH scenarios on Defects4J V2.0 are denoted as `-', as their original studies only evaluate the SL scenario for this version. We adhere to their original experimental settings during our reproduction.

\textbf{Robustness in Complex Scenarios.} As shown in Table~\ref{tab:result_three_repair_scenarios}, CausalRepair demonstrates a substantial advantage in the most challenging single-function scenario. Under the unified experimental setting using DeepSeek-V3, CausalRepair fixes 162 and 151 SF bugs on Defects4J V1.2 and V2.0, respectively. In comparison, the second best baseline ReinFix fixes only 118 and 118 SF bugs on the two versions, respectively, while TSAPR fixes 108 and 116 bugs, respectively. A similar performance gap is observed in the single-hunk scenario. Specifically, on Defects4J V1.2, CausalRepair fixes 109 SH bugs, outperforming TSAPR (87 bugs) and ReinFix (83 bugs). On Defects4J V2.0, CausalRepair maintains this lead by fixing 115 SH bugs, whereas both ReinFix and TSAPR fix only 89 bugs. This indicates that the causal context constructed through dual slicing enables CausalRepair to effectively filter out unexecuted code, helping the LLM maintain a precise focus on the root cause in complex repair scenarios.

\begin{table}[h]
\scriptsize
\caption{Repair results (correct fixes / plausible fixes) of different repair scenarios for CausalRepair and baselines on Defects4J.}
\label{tab:result_three_repair_scenarios} 
% \resizebox{1.0\columnwidth}{!}{
\begin{tabular}{@{}cc|ccc|ccc|c@{}}
\toprule
                              &                                      & \multicolumn{3}{c|}{Defects4J V1.2}                                                                                          & \multicolumn{3}{c|}{Defects4J V2.0}                                                                                          &                                          \\ \cmidrule(lr){3-8}
\multirow{-2}{*}{LLM}         & \multirow{-2}{*}{APR Method}         & \# SF Bugs                               & \# SH Bugs                               & \# SL Bugs                             & \# SF Bugs                               & \# SH Bugs                               & \# SL Bugs                             & \multirow{-2}{*}{Total}                  \\ \midrule
                              & BaseDeepSeek                         & 78/105                                   & 57/72                                    & 36/45                                  & 68/83                                    & 54/67                                    & 30/38                                  & 146/188                                  \\
                              & ChatRepair                           & 103/156                                  & 51/67                                    & 45/56                                  & -                                        & -                                        & 42/53                                  & 145/209                                  \\
                              & ThinkRepair                          & 87/122                                   & 66/89                                    & 38/48                                  & 93/129                                   & 68/98                                    & 35/51                                  & 180/251                                  \\
                              & ContrastRepair                       & 133/182                                  & 86/124                                   & 60/69                                  & -                                        & -                                        & 48/66                                  & 181/248                                  \\
                              & RepairAgent                          & 85/93                                    & 67/73                                    & 48/52                                  & 87/94                                    & 72/78                                    & 45/48                                  & 172/187                                  \\
                              &  AdverIntent-Agent                          &  84/93                                    &  66/74                                    &  47/51                                  &  87/96                                    &  70/77                                    &  46/51                                  &  171/189                                  \\
                              & TSAPR                                & 108/146                                  & 87/115                                   & 50/66                                  & 116/150                                  & 89/114                                   & 46/63                                  & 224/296                                  \\
                              & ReinFix                              & 118/181                                  & 83/120                                   & 47/66                                  & 118/171                                  & 89/128                                   & 47/63                                  & 236/352                                  \\
\multirow{-8}{*}{DeepSeek-V3} & \cellcolor[HTML]{EFEFEF}CausalRepair & \cellcolor[HTML]{EFEFEF}\textbf{162/191} & \cellcolor[HTML]{EFEFEF}\textbf{109/123} & \cellcolor[HTML]{EFEFEF}\textbf{62/69} & \cellcolor[HTML]{EFEFEF}\textbf{151/174} & \cellcolor[HTML]{EFEFEF}\textbf{115/128} & \cellcolor[HTML]{EFEFEF}\textbf{57/63} & \cellcolor[HTML]{EFEFEF}\textbf{313/365} \\ \midrule
                              & ReinFix                              & 124/-                                    & 93/-                                     & 57/-                                   & 130/-                                    & 103/-                                    & \textbf{56/-}                                   & 254/-                                    \\
\multirow{-2}{*}{GPT-4o}      & \cellcolor[HTML]{EFEFEF}CausalRepair & \cellcolor[HTML]{EFEFEF}\textbf{152/184} & \cellcolor[HTML]{EFEFEF}\textbf{105/122} & \cellcolor[HTML]{EFEFEF}\textbf{62/70} & \cellcolor[HTML]{EFEFEF}\textbf{145/175} & \cellcolor[HTML]{EFEFEF}\textbf{109/131} & \cellcolor[HTML]{EFEFEF}53/60 & \cellcolor[HTML]{EFEFEF}\textbf{297/359} \\ \bottomrule
\end{tabular}
% }
% \vspace{-0.1in}
\end{table}

\textbf{Effectiveness in Simple Scenarios.} For the relatively simple single-line bugs, CausalRepair also maintains its leading performance. On Defects4J V1.2 and V2.0, CausalRepair correctly fixes 62 and 57 SL bugs, respectively, outperforming ReinFix, which fixes 47 bugs on both versions, and ContrastRepair, which fixes 51 and 48 bugs, respectively. These results suggest that even for simple single-line repairs, the purified test context and the source context containing runtime traces provided by CausalRepair offer effective guidance, thereby achieving comprehensive superiority across different repair scenarios.

\subsection{RQ3: Ablation Study}

% To evaluate the contribution of each component in CausalRepair, we design two groups of variants for ablation studies, with the results summarized in Table~\ref{tab:result_ablation}.

{To evaluate the contribution of each core component in CausalRepair across its three phases, we construct five variants for ablation studies. The results are summarized in Table~\ref{tab:result_ablation}.}

% To evaluate the contribution of the construction of minimal causal context in CausalRepair, we construct three variants for ablation studies, with the results summarized in Table~\ref{tab:result_ablation}.

% \textbf{Variant Settings.} First, to investigate the necessity of causal context construction, we design three context slicing variants. 1) \textit{w/o Test Static Slicing (Full)} replaces the purified test context with the full test function to examine the impact of noise interference. 2) \textit{w/o Test Static Slicing (Line)} provides only the failing line and error message to assess the impact of dependency absence. 3) \textit{w/o Source Dynamic Slicing} removes the dynamic slicing information on the source side while keeping other components unchanged. Second, to evaluate the benefits of the repair strategies, we introduce two strategy variants. \textit{w/o Iterative Repair} disables the multi-turn conversational repair, and \textit{w/o Patch Augmentation} disables the subsequent patch augmentation.

% \textbf{Variant Settings.} We design three context variants to investigate the necessity of different context components: 1) \textit{w/o Source Dynamic Slicing} removes the dynamic slicing information on the source side while keeping other components unchanged. 2) \textit{w/o Test Static Slicing (Full)} replaces the purified test context with the full test function to examine the impact of noise interference. 3) \textit{w/o Test Static Slicing (Line)} provides only the failing line and error message to assess the impact of dependency absence.

\textbf{Variant Settings.} First, to investigate the necessity of different context components, we design three context variants: 1) \textit{w/o Source Dynamic Slicing} removes the dynamic slicing information on the source side while keeping other components unchanged. 2) \textit{w/o Test Static Slicing (Full)} replaces the purified test context with the full test function to examine the impact of noise interference. 3) \textit{w/o Test Static Slicing (Line)} provides only the failing line and error message to assess the impact of dependency absence. 
{Second, to evaluate the effectiveness of our multi-stage repair pipeline, we additionally introduce two variants: 4) \textit{w/o Iterative Repair} disables the conversation-driven repair loop, forcing the LLM to generate patches in a single-shot manner. 5) \textit{w/o Patch Augmentation} disables the final patch augmentation phase, outputting the first plausible patch found without exploring alternative implementations.}

\textbf{Impact of Source Context.} As shown in Table~\ref{tab:result_ablation}, removing source-side dynamic slicing substantially degrades repair performance, reducing the number of correct fixes by 26.5\%. This result shows that runtime execution traces provide critical causal evidence for repair. Without such source-side context, the LLM lacks explicit guidance on how the failure propagates through the executed code, making root-cause identification more difficult.

\textbf{Impact of Test Context.} Test-side slicing is also crucial. \textit{w/o Test Static Slicing (Line)} suffers the largest drop of 38.3\%, indicating that using only the failing line and error message causes severe dependency absence and prevents the LLM from understanding the test intent. In contrast, \textit{w/o Test Static Slicing (Full)} drops by 10.0\%, showing that full-test context introduces redundant noise. These results demonstrate that context-aware test slicing strikes a better balance by preserving failure-relevant dependencies while filtering irrelevant test scenarios.

{
\textbf{Impact of Iterative Repair and Patch Augmentation.} 
Beyond context construction, the subsequent repair phases also play crucial roles. Disabling the iterative repair loop (\textit{w/o Iterative Repair}) results in an 18.5\% drop in correct fixes (from 313 to 255), highlighting that without closed-loop validation feedback, the LLM cannot progressively refine its initial generated incorrect patch. Furthermore, disabling the final patch augmentation phase (\textit{w/o Patch Augmentation}) leads to a 16.9\% decrease in correct fixes (from 313 to 260). Due to test suite incompleteness, the initial plausible patch generated is often not semantically correct. By using this initial plausible patch as a reference to explore the surrounding solution space, the augmentation phase effectively increases the likelihood of generating a correct patch.
}

\begin{table}[h]
\scriptsize
\caption{Ablation results (correct fixes / plausible fixes) of CausalRepair with different components.}
\label{tab:result_ablation} 
\vspace{-0.1in}
% \resizebox{1.0\columnwidth}{!}{
\begin{tabular}{@{}lcccc@{}}
\toprule
{\color[HTML]{1F1F1F} Variants}                       & {\color[HTML]{1F1F1F} Defects4J V1.2}     & {\color[HTML]{1F1F1F} Defects4J V2.0}     & {\color[HTML]{1F1F1F} Total}            & {\color[HTML]{1F1F1F} Drop Rate}           \\ \midrule
% {\color[HTML]{1F1F1F} \textit{Context Variants}}  & {\color[HTML]{1F1F1F} }                   & {\color[HTML]{1F1F1F} }                   & {\color[HTML]{1F1F1F} }                 & {\color[HTML]{1F1F1F} }                    \\
{\color[HTML]{1F1F1F} w/o Source Dynamic Slicing}         & {\color[HTML]{1F1F1F} 106/142}            & {\color[HTML]{1F1F1F} 124/154}            & {\color[HTML]{1F1F1F} 230/296}          & {\color[HTML]{1F1F1F} $\downarrow$ 26.5\%} \\
{\color[HTML]{1F1F1F} w/o Test Static Slicing (Full)}    & {\color[HTML]{1F1F1F} 144/167}            & {\color[HTML]{1F1F1F} 138/162}            & {\color[HTML]{1F1F1F} 282/329}          & {\color[HTML]{1F1F1F} $\downarrow$ 10.0\%} \\
{\color[HTML]{1F1F1F} w/o Test Static Slicing (Line)}    & {\color[HTML]{1F1F1F} 88/106}             & {\color[HTML]{1F1F1F} 105/118}            & {\color[HTML]{1F1F1F} 193/224}          & {\color[HTML]{1F1F1F} $\downarrow$ 38.3\%} \\ 
% {\color[HTML]{1F1F1F} \textit{Strategy Variants}} & {\color[HTML]{1F1F1F} }                   & {\color[HTML]{1F1F1F} }                   & {\color[HTML]{1F1F1F} }                 & {\color[HTML]{1F1F1F} }                    \\
{\color[HTML]{1F1F1F} w/o Iterative Repair}       & {\color[HTML]{1F1F1F} 131/151}            & {\color[HTML]{1F1F1F} 124/140}            & {\color[HTML]{1F1F1F} 255/291}          & {\color[HTML]{1F1F1F} $\downarrow$ 18.5\%} \\
{\color[HTML]{1F1F1F} w/o Patch Augmentation}     & {\color[HTML]{1F1F1F} 131/191}            & {\color[HTML]{1F1F1F} 129/174}            & {\color[HTML]{1F1F1F} 260/365}          & {\color[HTML]{1F1F1F} $\downarrow$ 16.9\%} \\ \midrule
% w/o Iter. + Aug. (CausalRepair-Base)         & 110/151                                   & 110/140                                   & 220/291                                 & {\color[HTML]{1F1F1F} $\downarrow$ 36.1\%} \\ \midrule
{\color[HTML]{1F1F1F} \textbf{CausalRepair}}      & {\color[HTML]{1F1F1F} \textbf{162 / 191}} & {\color[HTML]{1F1F1F} \textbf{151 / 174}} & {\color[HTML]{1F1F1F} \textbf{313/365}} & {\color[HTML]{1F1F1F} \textbf{-}}          \\ \bottomrule
\end{tabular}
% }
% \vspace{-0.1in}
\end{table}

% \textbf{Impact of Repair Strategies and Intrinsic Effectiveness.} Regarding repair strategies, the results for \textit{w/o Iterative Repair} and \textit{w/o Patch Augmentation} show that the iterative mechanism and the augmentation mechanism further improve repair effectiveness, contributing 18.5\% and 16.9\% to the performance, respectively. However, an important concern arises as to whether the superiority of CausalRepair merely stems from these general strategy enhancements. To address this concern and fairly compare the intrinsic quality of the provided context, we further introduce a base variant that removes both iterative repair and patch augmentation, referred to as CausalRepair-Base. This variant performs one-shot generation with a patch size of five, which is consistent with the generation setting of ReinFix \cite{zhang2025repair}. Experimental results show that CausalRepair-Base fixes 220 bugs on Defects4J. For a fair comparison, we report the performance of ReinFix when generating the top-5 patches, which repairs only 200 bugs. This comparison (220 vs. 200) demonstrates that the core advantage of CausalRepair does not stem from increasing the number of conversation turns or sampling budget, but rather from the higher quality of its causal context. Even under a constrained generation budget, the precise causal evidence provided by CausalRepair enables the LLM to generate correct patches more effectively.

\begin{figure}[htbp]
    \centering
    \vspace{-0.1in}
    {\includegraphics[width=0.8\linewidth]{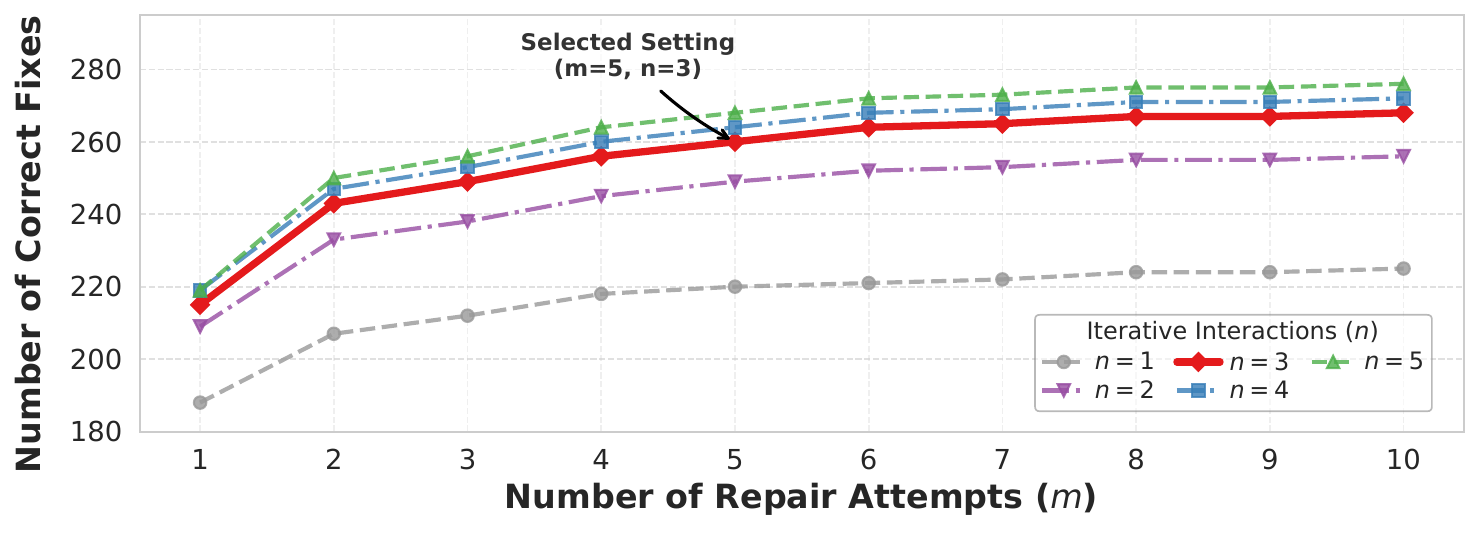}}
    \vspace{-0.1in}
    \caption{Impact of hyperparameter settings ($m$ and $n$) on repair effectiveness.} 
    \label{fig:hyperparameter}
    \vspace{-0.1in}
\end{figure}

\textbf{Parameter Sensitivity.} We further investigate the impact of the repair budget settings, i.e., the number of repair attempts ($m$) and iterative interactions ($n$), on CausalRepair's effectiveness. 
% Figure~\ref{fig:hyperparameter} illustrates the number of correct fixes under varying settings. 
{Figure~\ref{fig:hyperparameter} illustrates the number of correct fixes under varying repair budget settings of Phase 2, where Phase 3 is disabled to isolate the impact of $m$ and $n$ without the additional cost of patch augmentation.}
As shown in Figure~\ref{fig:hyperparameter}, we observe that the repair performance generally improves as $m$ and $n$ increase, as a larger search space allows the LLM to explore more potential solutions. However, the performance gain diminishes and saturates after $m=5$ and $n=3$. Considering the trade-off between repair effectiveness and computational cost (e.g., time and token consumption), we identify $m=5$ and $n=3$ as the optimal configuration for our main experiments.

\subsection{RQ4: Data Contamination Analysis}

{Considering the potential risk that LLMs may memorize training data, we conduct additional experiments on the recently released Defects4J-Trans \cite{li2025evaluating} benchmark to mitigate potential data leakage risks and to evaluate the robustness of CausalRepair against data memorization.} This dataset is constructed by applying semantic-preserving code transformations to the original bugs in Defects4J, and its release date (October 2024) is later than the training data cutoff of DeepSeek-V3. Therefore, Defects4J-Trans further reduces the risk of data leakage during the repair process. Under this experimental setting, we compare CausalRepair with BaseDeepSeek as well as the current state-of-the-art LLM-based approach ReinFix \cite{zhang2025repair}.

As shown in Table~\ref{tab:result_d4jtrans}, CausalRepair repairs a total of 289 bugs on Defects4J-Trans, significantly outperforming ReinFix with 200 repaired bugs and BaseDeepSeek with 97 repaired bugs. Specifically, CausalRepair repairs 144 and 145 bugs on versions V1.2 and V2.0, respectively, achieving the best repair performance on both versions. 
% These results highlight the strong generalization capability of CausalRepair, demonstrating that it not only performs well on Defects4J but also effectively generates high-quality patches when applied to unseen and transformed bug scenarios.
{These results highlight CausalRepair's robustness against data contamination, suggesting that its effectiveness is unlikely to be solely attributable to memorization.}

\begin{table}[h]
\scriptsize
% \begin{threeparttable}
\caption{Repair results (correct fixes / plausible fixes) for CausalRepair and baselines on Defects4J-Trans.}
\label{tab:result_d4jtrans} 
\vspace{-0.1in}
% \resizebox{1.0\columnwidth}{!}{
\begin{tabular}{@{}cc|ccc|ccc|c@{}}
\toprule
                             &                              & \multicolumn{3}{c|}{Defects4J-Trans V1.2}            & \multicolumn{3}{c|}{Defects4J-Trans V2.0}            &                         \\ \cmidrule(lr){3-8}
\multirow{-2}{*}{APR Method} & \multirow{-2}{*}{Patch Size} & \# SF Bugs       & \# SH Bugs       & \# SL Bugs     & \# SF Bugs       & \# SH Bugs       & \# SL Bugs     & \multirow{-2}{*}{Total} \\ \midrule
BaseDeepSeek                      & 25                           & 50/84          & 38/57           & 28/37          & 47/81           & 35/58           & 18/30          & 97/165                 \\
ReinFix                      & 45                           & 102/159          & 71/110           & 39/63          & 98/141           & 74/104           & 34/49          & 200/300                 \\
\rowcolor[HTML]{EFEFEF} 
CausalRepair                 & 25                           & \textbf{144/163} & \textbf{106/117} & \textbf{61/68} & \textbf{145/161} & \textbf{106/119} & \textbf{49/58} & \textbf{289/324}        \\ \bottomrule
\end{tabular}
% }
% \begin{tablenotes}
% \footnotesize
% \item \textit{SF}: Single-Function, \textit{SH}: Single-Hunk, \textit{SL}: Single-Line.
% \item “Correct fixes” are patches passing all ground-truth tests; “plausible fixes” pass all available tests.
% \end{tablenotes}

% \end{threeparttable}
\vspace{-0.1in}
\end{table}

\subsection{RQ5: Cost Analysis}
To evaluate the cost of CausalRepair, we measure the average repair time (including preprocessing, LLM inference, and validation) and monetary costs of various LLM-based methods on Defects4J, as presented in Table~\ref{tab:result_cost}. The results demonstrate that CausalRepair maintains the lowest time and monetary costs while achieving the best repair effectiveness. Specifically, CausalRepair requires only 23.28 minutes (including $\sim$2.53 minutes for dynamic slicing) and \$0.029 on average to fix a bug. Compared to the current state-of-the-art method ReinFix, the time overhead of CausalRepair is only 49.1\% of that of ReinFix, and the monetary cost is merely 17.9\%. Furthermore, although the monetary cost of CausalRepair is on par with TSAPR (both at \$0.029), our repair time is substantially shorter, with a larger number of fixed bugs.

% \begin{table}[h]
% \scriptsize
% % \begin{threeparttable}
% \caption{Cost analysis for CausalRepair and baselines on Defects4J.}
% \label{tab:result_cost} 
% % \resizebox{1.0\columnwidth}{!}{
% \begin{tabular}{@{}cccccccc@{}}
% \toprule
% Method     & ChatRepair & ThinkRepair & ContrastRepair & RepairAgent & TSAPR            & ReinFix   & CausalRepair       \\ \midrule
% Patch Size & 500        & 125         & 160            & 160         & 32               & 45        & \textbf{25}        \\
% Time/Bug   & 363.04 min & 326.29 min  & 251.17 min     & 139.53 min  & 42.32 min        & 47.42 min & \textbf{23.28 min} \\
% Money/Bug  & \$0.745    & \$0.438     & \$0.103        & \$0.429     & \textbf{\$0.029} & \$0.162   & \textbf{\$0.029}   \\ \bottomrule
% \end{tabular}
% % }
% \vspace{-0.1in}
% \end{table}

\begin{table}[h]
\scriptsize
% \begin{threeparttable}
\caption{Cost analysis for CausalRepair and baselines on Defects4J.}
\label{tab:result_cost} 
\vspace{-0.1in}
\resizebox{1.0\columnwidth}{!}{
\begin{tabular}{@{}ccccccccc@{}}
\toprule
Method     & ChatRepair & ThinkRepair & ContrastRepair & RepairAgent &  AdverIntent-Agent & TSAPR            & ReinFix   & CausalRepair       \\ \midrule
Patch Size & 500        & 125         & 160            & 160         &  36                & 32               & 45        & \textbf{25}        \\
Time/Bug   & 363.04 min & 326.29 min  & 251.17 min     & 139.53 min  &  78.76 min         & 42.32 min        & 47.42 min & \textbf{23.28 min} \\
Money/Bug  & \$0.745    & \$0.438     & \$0.103        & \$0.429     &  \$0.387           & \textbf{\$0.029} & \$0.162   & \textbf{\$0.029}   \\ \bottomrule
\end{tabular}
}
\vspace{-0.1in}
\end{table}

\section{Discussion}

\subsection{Evaluation on Recent Real-World Benchmarks}
\label{sec:discussion-recent-benchmarks}

% {Although Defects4J is the most widely used benchmark in APR, relying solely on Defects4J may limit the external validity of the evaluation. To further assess the practical applicability and generalizability of CausalRepair, we conduct additional experiments on two recent real-world benchmarks, i.e., RWB and GitBug-Java.}

{Although Defects4J is the most widely used benchmark in APR, relying solely on Defects4J may limit the external validity of the evaluation. 
To further assess the practical applicability and generalizability of CausalRepair, we conduct additional experiments on two recent real-world benchmarks, i.e., RWB and GitBug-Java, which have been widely used to evaluate the generalization capability of APR methods \cite{yin2024thinkrepair, zhang2025repair, bouzenia2025repairagent, silva2025repairllama}.}

\begin{itemize}
    \item \textbf{{RWB.}}
    {RWB (Real-World Bugs) was constructed by Yin et al.~\cite{yin2024thinkrepair}, a benchmark for evaluating recent real-world bugs collected from high-quality open-source projects. RWB contains two versions: RWB V1.0 consists of 44 SF bugs collected from bug-fixing commits after October 2021, and RWB V2.0 consists of 29 SF bugs collected after March 2023.}
    \item \textbf{{GitBug-Java.}}
    {GitBug-Java \cite{silva2024gitbug} is a reproducible benchmark of 199 recent Java bugs collected from the 2023 commit histories of 55 open-source repositories. GitBug-Java contains more diverse and complex repair scenarios. Among the 199 bugs, 132 are SF bugs, 47 are multi-function (MF) bugs, and the remaining 20 involve buggy code elements outside functions (Other Bugs). Additionally, 17 of these bugs require cross-file modifications, indicating that GitBug-Java includes repair tasks whose fixes span broader program scopes. Therefore, GitBug-Java provides a more realistic benchmark for evaluating repair capability beyond curated SF settings.}
\end{itemize}

{All additional experiments are conducted using the same DeepSeek-V3 backbone and configurations as our main evaluation.
Table~\ref{tab:rwb_results} reports the results on RWB. CausalRepair correctly fixes 31 out of 44 bugs on RWB V1.0 and 21 out of 29 bugs on RWB V2.0, achieving repair rates of 70.5\% and 72.4\%, respectively. 
Table~\ref{tab:recent_gitbug} further presents the results on GitBug-Java. CausalRepair correctly fixes 101 out of 199 bugs. For SF bugs, it fixes 73 out of 132 bugs, achieving a repair rate of 55.3\%. 
% Notably, CausalRepair also fixes 21 out of 47 MF bugs, showing that the proposed dual-slicing strategy can still provide useful causal evidence for more complex real-world bugs whose developer patches may span multiple functions and files.
% Overall, these results show that CausalRepair generalizes beyond Defects4J to recent, realistic repair benchmarks.
Notably, CausalRepair also fixes 21 out of 47 MF bugs and 7 out of 20 Other bugs, indicating that the proposed dual-slicing strategy remains applicable when the repair scope extends beyond a single function.
Overall, these results demonstrate the effectiveness of CausalRepair on recent and realistic repair benchmarks beyond Defects4J.}

\begin{table}[h]
\scriptsize
\centering
\caption{Repair Performance of CausalRepair on RWB.}
\vspace{-0.1in}
\label{tab:rwb_results}
\begin{tabular}{lcccc}
\toprule
\multirow{2}{*}{Project} & \multicolumn{2}{c}{RWB V1.0} & \multicolumn{2}{c}{RWB V2.0} \\
\cmidrule(lr){2-3} \cmidrule(lr){4-5}
& \# Correct Fixes & \# Bugs & \# Correct Fixes & \# Bugs \\
\midrule
Cli         & 4  & 4  & 4  & 4  \\
Codec       & 4  & 5  & 4  & 4  \\
Collections & 1  & 1  & -  & -  \\
Compress    & 5  & 8  & 2  & 4  \\
Csv         & 1  & 2  & -  & -  \\
Jsoup       & 12 & 18 & 7  & 11 \\
Lang        & 4  & 6  & 4  & 6  \\
\midrule
Total (Rate) & 31 (70.5\%) & 44 & 21 (72.4\%) & 29 \\
% Repair Rate & 70.5\% & - & 72.4\% & - \\
\bottomrule
\end{tabular}
\vspace{-0.2in}
\end{table}
\begin{table}[h]
\scriptsize
\centering
\caption{Repair performance of CausalRepair on GitBug-Java.}
\vspace{-0.1in}
\label{tab:recent_gitbug}
\begin{tabular}{lccc}
\toprule
Bug Type & \# Bugs & \# Correct Fixes & Repair Rate \\
\midrule
\# SF Bugs & 132 & 73 & 55.3\% \\
\# MF Bugs  & 47  & 21 & 44.7\% \\
\# Other Bugs               & 20  & 7  & 35.0\%  \\
\midrule
Total                & 199 & 101 & 50.8\% \\
\bottomrule
\end{tabular}
\vspace{-0.1in}
\end{table}

\subsection{Robustness under Realistic Fault Localization}
\label{sec:discussion-realistic-fl}

{In the preceding RQs, we evaluate CausalRepair under the assumption of perfect FL, which is widely adopted in the APR community to ensure a controlled environment for comparing isolated patch generation capabilities.
Nevertheless, in practical APR scenarios, the FL is usually provided by automated FL tools, which are inherently imprecise. To mitigate this threat to validity and assess the practical applicability of CausalRepair, we further evaluate its performance under a realistic FL setting on Defects4J.

Following prior studies \cite{ye2024iter, liu2019tbar, jiang2018shaping, li2020dlfix, liu2018lsrepair, wen2018context, xuan2016nopol}, we utilize Spectrum-Based Fault Localization (SBFL). Specifically, we employ the widely used GZoltar \cite{campos2012gzoltar} framework alongside the Ochiai \cite{abreu2007accuracy} formula to compute the suspiciousness score for each statement. Following \cite{ye2024iter}, we consider the top-50 most suspicious statements as repair candidates for each bug. For each candidate, we supply its enclosing function to CausalRepair to construct the causal context by dual-slicing and execute the repair process. We utilize the same DeepSeek-V3 backbone as in our main experiments, and the remaining repair configuration is kept the same.
}

% {Table~\ref{tab:realistic_fl} presents the repair performance of CausalRepair under the realistic FL setting, alongside the top-performing baselines evaluated under perfect FL. Under this setting, CausalRepair correctly fixes 232 bugs on Defects4J, which preserves 74.1\% of its effectiveness under perfect FL. Although the performance inevitably decreases compared with the perfect FL setting, this drop is mainly attributable to SBFL imprecision: when the top-50 suspicious lines do not cover the actual buggy function, CausalRepair is forced to repair a non-buggy function, making the correct patch unreachable regardless of the quality of the constructed context. Nevertheless, CausalRepair still outperforms most LLM-based baselines evaluated under perfect FL and achieves neck-to-neck performance with the strongest baseline ReinFix. This result indicates that CausalRepair remains effective even when the initial FL is provided by an imperfect FL tool.

{
Under this realistic FL setting, CausalRepair correctly fixes 232 bugs on Defects4J, showing that it remains effective when the initial FL is produced by an imperfect automated FL tool. 
This effectiveness stems from the dual-slicing design: once a suspicious candidate covers failure-relevant code, CausalRepair can refine the repair context by preserving relevant test semantics and executed dependencies, rather than relying solely on the initial suspicious statement. 
Overall, these results are promising and suggest that CausalRepair is robust to noisy localization signals and remains practically useful under more realistic APR settings.
}

% {Under this realistic FL setting, CausalRepair correctly fixes 232 bugs on Defects4J. This result demonstrates that CausalRepair can still maintain strong repair capability when the initial FL is provided by an imperfect FL tool. This robustness benefits from the dual-slicing design: when a suspicious candidate covers failure-relevant code, CausalRepair further refines the repair context by preserving the relevant test semantics and executed dependencies, rather than directly relying on the initial suspicious statement. Notably, even under this more challenging setting, CausalRepair still fixes more bugs than most LLM-based baselines evaluated under perfect FL in the preceding RQs, and achieves neck-to-neck performance with the strongest baseline ReinFix. These findings further suggest that CausalRepair remains effective in more realistic APR scenarios and improves the practical applicability of LLM-based APR.}

% \vspace{-0.1in}
\section{Threats to Validity}

\textbf{Construct Validity.} The primary threat to construct validity lies in the subjectivity of patch correctness assessment. 
Since a plausible fix is not necessarily semantically correct, we rely on manual inspection to identify correct patches. To mitigate this threat, we follow prior work \cite{xia2023automated, xia2024automated, ye2022selfapr} and adopt a rigorous validation mechanism. Specifically, each patch is independently reviewed by two experienced Java researchers for semantic equivalence to the developer patch. Disagreements are resolved by involving a third developer until consensus is reached.
% , improving objectivity and accuracy of the evaluation results. 
% Since plausible fixes are not necessarily semantically correct, we manually validate patch correctness following prior work \cite{xia2023automated, xia2024automated, ye2022selfapr}. Each patch is independently reviewed by two experienced Java researchers for semantic equivalence to the developer patch. Disagreements are resolved by involving a third developer until consensus is reached.
Furthermore, we have open-sourced our generated patches for public assessment \cite{CausalRe59:online}.

% \noindent\textbf{Internal Validity.} The main threat to internal validity is the risk of data leakage; that is some developer-generated patches may overlap with the training data used by LLMs. To mitigate this threat, in addition to the widely used Defects4J dataset, we evaluate CausalRepair on an additional dataset, Defects4J-Trans, which is less likely to be affected by data contamination. The results show that CausalRepair’s effectiveness on this dataset is comparable to that achieved on Defects4J, suggesting that the experimental results achieved by CausalRepair on Defects4J are not significantly affected by data contamination issues.
\noindent\textbf{Internal Validity.} The main threat to internal validity is the risk of data leakage; that is, some developer-generated patches may overlap with the training data used by LLMs. To mitigate this threat, in addition to the widely used Defects4J dataset, we evaluate CausalRepair on an additional dataset, Defects4J-Trans, which is less likely to be affected by data contamination. Additionally, all experiments are conducted under a unified LLM setting to ensure a fair comparison among different LLM-based methods. Under this controlled setting, CausalRepair consistently achieves the best performance across both datasets. These results suggest that the effectiveness of CausalRepair is unlikely to be significantly affected by data contamination issues.

% \noindent\textbf{External Validity.} External validity mainly concerns the generalizability of our approach. A potential limitation of our method is that its current implementation targets only the Java programming language (PL), primarily because the dynamic slicing tool we employ, \textit{Slicer4J}, is designed for the Java ecosystem. Focusing on a single PL may limit the general applicability of our findings to other PLs. However, as one of the most widely used PLs in both industry and academia, Java is highly representative. Furthermore, we conduct extensive evaluations on the widely recognized APR benchmark Defects4J as well as the additional Defects4J-Trans dataset, covering diverse types of real-world bugs. Therefore, our experimental results are still sufficient to demonstrate the effectiveness of CausalRepair. Additionally, our proposed dual-slicing paradigm is PL-agnostic and can be adapted to other PLs. We plan to extend our approach to other PLs in future work.

\noindent\textbf{External Validity.} External validity mainly concerns the generalizability of our approach. First, a potential limitation is that our current implementation targets only Java programming language (PL), primarily because the dynamic slicing tool we employ, \textit{Slicer4J}, is designed for the Java ecosystem. However, as one of the most widely used PLs in both industry and academia, Java is highly representative. Moreover, our proposed dual-slicing paradigm is PL-agnostic and can be adapted to other languages, which we plan to explore in future work. {Second, regarding dataset generalizability, we conduct extensive evaluations across multiple datasets. Beyond the widely recognized Defects4J and its transformed variant Defects4J-Trans, we incorporate recent real-world benchmarks including RWB and GitBug-Java. These diverse datasets encompass a wide range of realistic repair scenarios. Therefore, our comprehensive evaluation shows the broad applicability and effectiveness of CausalRepair.}

\section{Related Work}
\subsection{Automated Program Repair}

Automated Program Repair (APR) aims to assist developers in localizing and fixing program bugs automatically. Existing APR techniques can be broadly classified into traditional APR approaches, learning-based APR approaches, and Large Language Model (LLM)-based APR approaches.

\textbf{Traditional APR Approaches.}
Traditional APR approaches mainly fall into three categories: heuristic-based \cite{le2011genprog, jiang2018shaping, le2016history, wen2018context}, constraint-based \cite{demarco2014automatic, long2015staged, le2017s3, mechtaev2016angelix, gao2021beyond}, and template-based \cite{martinez2016astor, hua2018sketchfix, ghanbari2019practical, liu2019avatar, liu2019tbar}.
Heuristic-based approaches follow the generate-and-validate paradigm, exploring candidate patches guided by test cases. Constraint-based approaches formulate repair as a constraint satisfaction problem, extracting semantic constraints via symbolic execution and synthesizing patches with solvers. Template-based approaches, often regarded as the most effective traditional techniques, generate patches by matching predefined fix patterns.
However, traditional APR approaches have limited generality, as they are often tailored to specific bug types or constrained by a restricted patch search space, making it difficult to repair previously unseen or complex logical bugs.

\textbf{Learning-based APR Approaches.} With the rapid advancement of deep learning techniques, a variety of learning-based APR approaches have been proposed \cite{chen2019sequencer, jiang2021cure, li2020dlfix, li2022dear, lutellier2020coconut, drain2021deepdebug, meng2022improving, jiang2023knod, lin2024one}. These approaches introduce Neural Machine Translation (NMT) techniques and model program repair as a translation task from buggy code to correct code. For example, SequenceR \cite{chen2019sequencer} and CoCoNut \cite{lutellier2020coconut} leverage sequence-to-sequence models to learn repair patterns from large-scale code corpora. 
% To further improve syntactic correctness and compilation success rates, 
Subsequent work further incorporates domain-specific enhancements, such as syntax-guided decoding in Recoder \cite{zhu2021syntax} and execution-aware optimization in RewardRepair \cite{ye2022neural}.
% subsequent work has incorporated domain-specific enhancements. For example, Recoder \cite{zhu2021syntax} designs a syntax-guided edit decoder, while RewardRepair \cite{ye2022neural} proposes a loss function optimized using feedback from compilation and test execution. 
Although these methods significantly improve generalization capability, their performance remains constrained by the scale and diversity of training data, limiting their ability to handle unseen bug types.

\textbf{LLM-based APR Approaches.} Benefiting from the superior code understanding and generation capabilities of LLMs, LLM-based APR has emerged as the dominant paradigm in recent years. Early work such as AlphaRepair \cite{xia2022less} adopts a cloze-style generation scheme to exploit the zero-shot repair capability of LLMs. ChatRepair \cite{xia2024automated} introduces a conversation-driven paradigm, where test-execution feedback guides multi-round repair, and ContrastRepair \cite{kong2025contrastrepair} further incorporates contrastive test cases to improve conversational repair. To enrich repair context, approaches such as ReinFix \cite{zhang2025repair} and ThinkRepair \cite{yin2024thinkrepair} adopt retrieval-augmented strategies, using static analysis or vector-based retrieval to collect relevant definitions, method signatures, or repair templates. 
{Recent studies have also explored agentic workflows for APR. For example, RepairAgent \cite{bouzenia2025repairagent} and AdverIntent-Agent \cite{ye2025adverintent} formulate APR as an agent-driven process that analyzes failures, invokes tools, and iteratively refines patches. Other agentic systems, such as AutoCodeRover \cite{zhang2024autocoderover}, Agentless \cite{xia2024agentless}, and Prometheus \cite{chen2025prometheus}, target broader repository-level issue resolution on SWE-bench-style benchmarks. Unlike test-driven APR, these issue-resolution systems start from a natural-language issue description and the repository codebase, without assuming that an exposed failing test is available to the repair system; the official failing tests are hidden and used only for final evaluation. In contrast, CausalRepair focuses on test-driven APR, where the bug has already been reproduced by at least one failing test, and the failing test serves as an executable specification for constructing failure-specific causal repair context. Therefore, CausalRepair and SWE-bench-style issue-resolution agents are built upon different problem formulations, assumptions, and inputs.}
% {Recent studies have also explored agentic workflows for APR. RepairAgent \cite{bouzenia2025repairagent} and AdverIntent-Agent \cite{ye2025adverintent} formulate APR as an agent-driven process that analyzes failures, invokes tools, and iteratively refines patches. Other agentic systems, such as AutoCodeRover \cite{zhang2024autocoderover}, Agentless \cite{xia2024agentless}, and Prometheus \cite{chen2025prometheus}, target broader repository-level issue resolution on SWE-bench-style benchmarks, where the system starts from a natural-language issue description and may need to reproduce the bug and locate relevant code before generating patches, while final failing tests are hidden for validation. In contrast, CausalRepair focuses on test-driven APR after a concrete failure has been reproduced, and exploits the available failing test to construct failure-specific causal repair context. Thus, these issue-resolution agents operate under a different but potentially complementary setting, where CausalRepair could serve as a downstream repair component once an executable failing test is obtained.}

Despite the substantial progress of LLM-based APR, a causality gap in context construction remains. Test-side contexts either provide limited information or include redundant test scenarios, resulting in context ambiguity, while source-side contexts rely on static-analysis-based retrieval and introduce over-approximated dependencies. Consequently, the absence of precise runtime causal dependencies impedes accurate root cause localization, leading to incorrect patches. To bridge this gap, CausalRepair integrates precise runtime causal chains, captured via dual slicing, into a conversation-driven framework, preserving iterative feedback while providing LLMs with a high-signal causal context for more effective patch generation.

\subsection{Program Slicing}

Program slicing, originally formally defined by Mark Weiser \cite{weiser2009program}, is a semantics-based code decomposition technique designed to extract parts irrelevant to a specific point of interest (i.e., slicing criterion) from complex programs to construct a semantically consistent but smaller program subset. Based on the utilization of runtime information, slicing techniques are primarily categorized into static slicing and dynamic slicing \cite{tip1994survey, xu2005brief}. Static slicing analyzes all possible execution paths by constructing a Program Dependence Graph (PDG) or a System Dependence Graph (SDG) \cite{ferrante1987program, horwitz1990interprocedural}. Since static analysis must conservatively assume all possible branches and pointer aliases, it typically suffers from static over-approximation \cite{binkley2004survey}, resulting in slices containing a significant amount of redundant code that is not actually executed. In contrast, dynamic slicing relies on the execution trace of a specific input. It can precisely eliminate unexecuted paths and dead code, thereby capturing more accurate dependency relationships \cite{korel1988dynamic, agrawal1990dynamic}.

% Program slicing has been widely adopted throughout the software development lifecycle \cite{tip1994survey, xu2005brief}. In debugging and fault localization, dynamic slicing filters out code irrelevant to failure manifestation, substantially reducing the search space (e.g., by 60\%–90\%) and improving debugging efficiency \cite{gallagher2008program, demillo1996critical, lekssays2025llmxcpg}. In regression testing and impact analysis, forward slicing identifies program elements affected by code changes and their propagation paths, enabling precise impact analysis and efficient regression test selection \cite{gallagher2008program, gupta1996program, binkley1998application, acharya2011practical}. Furthermore, in the domain of security analysis, slicing is commonly realized as taint analysis to track malicious data flows and detect vulnerabilities such as SQL injection \cite{newsome2005dynamic, salimi2022vulslicer}. 
% Program slicing has been widely adopted throughout the software development lifecycle \cite{tip1994survey, xu2005brief}. It 
Program slicing has been widely applied across a range of software engineering tasks \cite{tip1994survey, xu2005brief}, including debugging and fault localization \cite{gallagher2008program, demillo1996critical, lekssays2025llmxcpg}, regression testing and impact analysis \cite{gallagher2008program, gupta1996program, binkley1998application, acharya2011practical}, as well as security analysis (e.g., taint analysis for vulnerability detection) \cite{newsome2005dynamic, salimi2022vulslicer, nong2025appatch}.
% With the rise of LLMs, program slicing has regained research attention due to its superior capability for semantic compression. It is primarily utilized to address the context window limit and the "lost-in-the-middle" \cite{liu2024lost} phenomenon faced by LLMs. Recent studies mainly leverage slicing to purify the input context for LLMs. 
With the rise of LLMs, program slicing has re-emerged as a key technique for semantically compressing input context, helping LLMs overcome context window limit and the “lost-in-the-middle” \cite{liu2024lost} phenomenon. For example, the RESCUE \cite{shi2025rescue} framework utilizes security-focused static program slicing to eliminate irrelevant code noise retrieved in Retrieval-Augmented Generation (RAG) systems, thereby enhancing the security of generated code. LLMxCPG \cite{lekssays2025llmxcpg} performs slicing using Code Property Graphs (CPG), reducing the code volume for vulnerability detection tasks by over 90\%. Additionally, ReduceFix \cite{yang2025input} draws on slicing concepts (specifically delta debugging) to automatically reduce error-inducing test inputs to fit within the LLM's context window.

However, existing works integrating LLMs are mostly limited to static slicing or input data reduction, leaving limitations in APR tasks. Existing retrieval-augmented APR approaches rely on static analysis for context retrieval. This inevitably introduces substantial unexecuted code and static noise, obscuring the causal chain triggering the error. Although dynamic slicing offers a high signal-to-noise ratio, its potential in LLM-based APR has not yet been fully exploited. 
% CausalRepair aims to bridge this gap. By first obtaining a purified test context via static slicing and then introducing dynamic slicing on the source code side, it filters out noisy code irrelevant to specific test failures, thereby constructing a causal context for the LLM.
CausalRepair bridges this gap by combining static slicing on the test side with dynamic slicing on the source code, filtering out code irrelevant to the failure and constructing a causal context for the LLM.

\section{Conclusion}

% In this paper, we introduce CausalRepair, a novel framework that employs a synergistic dual-slicing strategy to improve LLM-based conversational APR. CausalRepair performs context-aware static slicing on the test side to purify test semantics, and utilizes execution-trace-based dynamic slicing on the source side to construct a minimal causal context. Additionally, we incorporate a patch augmentation module to enhance patch diversity. Our experiments on Defects4J demonstrate that CausalRepair can fix a total of 313 bugs, which significantly outperforms other state-of-the-art baselines. We further demonstrate CausalRepair’s generalization ability on the unseen benchmark Defects4J-Trans. Compared to existing LLM-based APR tools, CausalRepair is more efficient and reduces the average repair cost to \$0.029 per bug.

In this paper, we first introduce the concept of minimal causal context. Based on this concept, we propose CausalRepair, a novel framework that instantiates this concept through a synergistic dual-slicing strategy. Specifically, CausalRepair performs context-aware static slicing on the test side to purify test semantics and utilizes execution-trace-based dynamic slicing on the source side, thereby constructing the minimal causal context to guide the iterative repair process. Our experiments on Defects4J demonstrate that CausalRepair can fix a total of 313 bugs, which significantly outperforms state-of-the-art baselines. Compared to existing LLM-based APR tools, CausalRepair is more efficient and reduces the average repair cost to \$0.029 per bug.

% \noindent{\textbf{Data Availability:}} 

\section{Data-Availability Statement}
To support reproducibility, the replication package and experimental results of this work are publicly available at: \href{https://doi.org/10.5281/zenodo.20953899}{https://doi.org/10.5281/zenodo.20953899} \cite{CausalRe59:online}.

\vspace{-0.1in}
\section*{Acknowledgments}

This work was supported by the National Key Research and Development Program of China under Grant No. 2024YFE0204200.

% \section{Acknowledgments}

\bibliographystyle{ACM-Reference-Format}
\bibliography{samples/sample-base-with-doi}

% \appendix

\end{document}